\documentclass[a4paper,12pt]{article}
\usepackage[english]{babel}
\usepackage[utf8]{inputenc}
\usepackage{t1enc}
\usepackage{floatflt}
\usepackage{graphicx}
\usepackage{psfrag}
\usepackage{bbm}
\usepackage{amsmath}
\usepackage{amssymb}
\usepackage{hyperref}
\usepackage{ifthen}
\usepackage{subcaption}
\usepackage{epstopdf}

\def\d{\mathrm{d}}
\def\e{\mathrm{e}}
\def\imagi{\mathrm{i}}

\def\kihagy#1{}

\newcommand{\arxiv}[2][]{%
  \ifthenelse{\equal{#1}{}}{%
    \href{http://arxiv.org/abs/#2}{\texttt{arXiv:#2}}%
  }{%
    \href{http://arxiv.org/abs/#2}{\texttt{arXiv:#2 [#1]}}%
  }%
}%

\newcommand{\doix}[2]{\href{http://dx.doi.org/#2}{#1}}%

\title{On non-topological solitons in Abelian gauge theories coupled to U(1)$\times$U(1) symmetric scalar fields}
\author{Péter Forgács\textsuperscript{1,2} and Árpád Lukács\textsuperscript{3,1}\\
{\small {}\textsuperscript{1} Wigner RCP RMI, POB 49, H1525 Budapest, Hungary.}\\
{\small {}\textsuperscript{2} Institut Denis-Poisson CNRS/UMR 7013, Universit\'e de Tours, Parc de Grandmont, 37200 Tours, France}\\
{\small {}\textsuperscript{3} Department of Theoretical Physics, University of the Basque Country UPV/EHU}\\
{\small POB 644, E-48080 Bilbao, Spain}
}

\begin{document}

\maketitle

\begin{abstract}
In a series of recent works Ishihara and Ogawa have investigated non-topological solitons (Q-balls) in a spontaneously broken Abelian gauge theory coupled to two complex scalar fields. The present paper extends their investigations to the most general U(1)$\times$U(1) symmetric quartic potential. Also, a new class of charged Q-ball solutions with vanishing self-interaction terms is investigated and some of their remarkable properties are exhibited.
\end{abstract}

The term Q-ball denotes finite energy, non-radiating solutions in theories containing scalar fields with time-periodic phases and associated conserved charges. Their discovery goes back to the pioneering works of G.~Rosen \cite{rosen}.  Prototype Q-balls appear in pure scalar field theories containing a complex scalar with a quartic potential coupled to a real one \cite{friedbergleesirlin}. They have been shown to be stable, their stability being related to their conserved charge. For an excellent review see Ref.\cite{LeePang}.
The term Q-ball comes from Ref.\ \cite{Coleman} where such scalar lumps with harmonic time dependent phases have been shown to occur in a scalar field theory containing a single complex scalar with a self-interaction potential of at least degree 6.
It turned out that similar lumps appear in gauge theories \cite{rosen2, Lee, BenciFortunato}, for some detailed numerical investigations see Refs.\ \cite{LeeYoon, GNS, GNPS} (with one complex, and one real scalar).
For recent reviews, see Refs.\ \cite{Shnir, NSrev}.

Q-balls have gained large attention due to the possibility of their formation in the early universe \cite{friemanetal}, them being candidates as dark matter \cite{kusenkoshaposhnikov}, their possible role in baryogenesis \cite{dodelsonwidrow}, and also their appearance in a large class of supersymmetric extensions of the Standard Model of Particle Physics \cite{kusenko,  enqvistmcdonald}.

The authors of Ref.\ \cite{ishiharaogawascreen} have found that for a spherically symmetric distribution of external charges coupled to an Abelian Higgs model, the Higgs field provides for perfect charge screening, cancelling out all long-range fields of the external charges. In Refs.\ \cite{ishiharaogawasoln, ishiharaogawasoln2}, this observation has been tested on Q-balls in an Abelian Higgs model coupled to another charged, massive scalar field whose mass is provided by the Higgs mechanism.
We note, that coupling the scalars this way is analogous to the much studied Higgs portal models \cite{SilveiraZee, PW} where a scalar dark sector is coupled exclusively to the Higgs fields of the Standard Model of Particle Physics.

In the present paper we extend the results of Refs.\ \cite{ishiharaogawasoln, ishiharaogawasoln2} to the case of
the most general U(1)$\times$ U(1) symmetric scalar sector with quartic self-interaction potentials.
We show that the remarkably precise numerically observed cancellation of the charge contribution between the two charged scalar fields pointed out in Refs.\ \cite{ishiharaogawasoln, ishiharaogawasoln2} follows from Gauss' theorem. It can therefore serve as an excellent test for the correctness of the numerical computations.
In carrying out a detailed investigation of a larger phase-space which appears to be a natural setting for the models considered, an interesting subfamily of charged Q-balls is found where the quartic self-interaction terms are put to zero. This new family of charged Q-balls is a natural extension of previously considered ungauged Q-balls with vanishing potential in Ref.\ \cite{LevRub} and investigated in more detail in Ref.\ \cite{LPSh}.

\section{The model considered}\label{sec:model}
Following Refs.\ \cite{ishiharaogawascreen, ishiharaogawasoln, ishiharaogawasoln2}, we shall consider an Abelian Higgs model containing two charged, complex scalar fields with an action given by
\begin{equation}\label{eq:act}
 S = \int \d^4 x \left[ -\frac{1}{4}F_{\mu\nu}F^{\mu\nu} + D_\mu\phi^* D^\mu\phi +D_\mu\psi^*D^\mu\psi- V\right]\,,
\end{equation}
where indices are raised and lowered by the Minkowski metric $g=\mathop{\text{diag}}(+,-,-,-)$, $F_{\mu\nu} = \partial_\mu A_\nu - \partial_{\mu}A_\nu$, $D_\mu \phi = (\partial_\mu -\imagi e_1A_\mu)\phi$, $D_\mu \psi = (\partial_\mu -\imagi e_2A_\mu)\psi$. The interaction potential is given as
\begin{equation}\label{eq:pot}
V = \frac{\lambda_1}{2}(|\phi|^2-\eta^2)^2 + \frac{\lambda_2}{2}|\psi|^4 +\lambda_{12}(|\phi|^2-\eta^2)|\psi|^2+m^2|\psi^2|\,,
\end{equation}
which is the most general quartic, gauge invariant potential for the two complex scalar fields with a U(1)$\times$ U(1) symmetry.

In the model defined by the action (\ref{eq:act}), there are two separately conserved U(1) currents:
\begin{equation}\label{eq:cons}
 j_{\phi\,\mu} = \imagi e_1(\phi^*D_\mu\phi - \phi D_\mu\phi^*)\,,
 \quad\quad
 j_{\psi\,\mu} = \imagi e_2(\psi^*D_\mu\psi - \psi D_\mu\psi^*)\,,
\end{equation}
The conserved charges are given as
\begin{equation}\label{eq:charges}
 Q_{\phi,\psi} = \int \d^3 x j_{\phi,\psi}^0\,.
\end{equation}



It is convenient to adimensionalise the fields and the coordinates as
$\phi\to\eta\phi$, $\psi\to\eta\psi$, $A_\mu \to \eta A_\mu$ and $x^\mu \to x^\mu/(e\eta)$ and in the action (\ref{eq:act}), which then becomes
\begin{equation}\label{eq:actR}
S = \frac{1}{e^2}\int\d^4 x\left[ -\frac{1}{4}F_{\mu\nu}F^{\mu\nu}+D_\mu \phi^*D^\mu\phi + D_\mu\psi^*D^\mu\psi - V\right]\,,
\end{equation}
with $q_i=e_i/e$, ($i=1,2$), $D_\mu\phi = (\partial_\mu - \imagi q_1A_\mu)\phi$, $D_\mu\psi = (\partial_\mu - \imagi q_2A_\mu)\psi$
\begin{equation}\label{eq:potR}
 V = \frac{\beta_1}{2}(|\phi|^2-1)^2 + \frac{\beta_2}{2}|\psi|^4 +\beta_{12}(|\phi|^2-1)|\psi|^2 + \mu |\psi|^2\,,
\end{equation}
where $\beta_{1,2} = \lambda_{1,2}/e^2$, $\beta_{12}=\lambda_{12}/e^2$, and $\mu = m^2/(e^2\eta^2)$.

\section{Spherically symmetric Q-ball solutions}\label{sec:Qballs}

We shall seek spherically symmetric Q-ball solutions of the model defined by the action (\ref{eq:act}).
The simplest spherically symmetric, purely electric Ansatz leading to non-trivial finite energy solutions can be written in the Lorenz gauge as
\begin{equation}\label{eq:Ansatz0}
 A_0 = {\tilde\alpha}(r)\,,\quad \phi = f_1(r)\e^{\imagi \omega_1 t}\,,\quad \psi=\e^{\imagi \omega_2 t} f_2(r)\,,
\end{equation}
where $r,\vartheta,\varphi$ are spherical coordinates, and all other vector potential components vanish.
It is convenient to gauge transform \eqref{eq:Ansatz0} to a simpler form where one of the scalars is time-independent, leading to the Ansatz also employed in Ref.\ \cite{ishiharaogawasoln, ishiharaogawasoln2}:
\begin{equation}\label{eq:Ansatz}
 A_0 = \alpha(r)\,,\quad \phi = f_1(r)\,,\quad \psi=\e^{\imagi \omega t} f_2(r)\,,
\end{equation}
which shall be also used in this paper.
The reduced action of the configuration within Ansatz (\ref{eq:Ansatz}) can be written as
\begin{equation}\label{eq:effact}
S_{\rm eff} =4\pi \int\d r r^2 (K_{\rm eff}-U_{\rm eff})\,,\ {\rm where}\
K_{\rm eff}=(f_1')^2+(f_2')^2-(\alpha')^2/2\,,
\end{equation}
with the effective potential given by
\begin{equation}\label{eq:effpot}
U_{\rm eff} = -\beta_1(f_1^2-1)^2/2-\beta_2 f_2^4/2-\beta_{12}(f_1^2-1)f_2^2-\mu f_2^2+q_1^2\alpha^2f_1^2+(q_2\alpha-\omega)^2 f_2^2\,.
\end{equation}
The spherically symmetric field equations resulting from the variation of the reduced action (\ref{eq:effact}) are given as
 \begin{align}
  \frac{1}{r^2}(r^2 f_1')' &= f_1 \left[ -q_1^2\alpha^2 +\beta_1(f_1^2-1)+\beta_{12} f_2^2\right]\,,\label{eq:radeq1}\\     \frac{1}{r^2}(r^2 f_2')' &= f_2 \left[ -(q_2\alpha-\omega)^2 +\beta_2 f_2^2+\mu+\beta_{12} (f_1^2-1)\right]\,,\label{eq:radeq2}\\
  \frac{1}{r^2}(r^2\alpha')' &= 2\left[q_1^2\alpha f_1^2 + q_2(q_2\alpha-\omega)f_2^2\right]\label{eq:radeq3}\,.
 \end{align}
The boundary conditions required for the solution of Eqs.\ (\ref{eq:radeq1}-\ref{eq:radeq3}) are derived, on one hand, from regularity of the fields at the origin,
\begin{equation}\label{eq:bdry0}
 f_1 \sim f_1(0) + f_1^{(2)}r^2+\dots\,,\quad f_2 \sim f_2(0) + f_2^{(2)}r^2+\dots\,,\quad \alpha \sim \alpha(0)+\alpha^{(2)}r^2+\dots\,,
\end{equation}
and from the requirement of vanishing energy density at infinity, i.e., approaching the vacuum manifold of the theory,
\begin{equation}\label{eq:bdryInf}
 f_1\to 1\,,\quad\quad f_2\to 0\,,\quad\quad \alpha\to 0\,.
\end{equation}

The energy of a field configuration defined by the Ansatz (\ref{eq:Ansatz})is expressed as
\begin{equation}\label{eq:erg}
 E = \frac{4\pi}{e}\eta\int_0^\infty \d r  r^2 \left[ (f_1')^2 + (f_2')^2 +\frac{1}{2}(\alpha')^2 + q_1^2\alpha^2 f_1^2 + (q_2\alpha-\omega)^2 f_2^2 + V\right]\,,
\end{equation}
where
\[V = \frac{\beta_1}{2}(f_1^2-1)^2+\frac{\beta_2}{2}f_2^4+\beta_{12}(f_1^2-1)f_2^2+\mu f_2^2\,.\]
The electric charges of $\phi$ resp.\ $\psi$ are written as
\begin{equation}\label{eq:chg}
 Q_\phi = \frac{4\pi}{e}\int \d r  r^2 \rho_\phi\,, \quad\quad Q_\psi = \frac{4\pi}{e}\int \d r r^2 \rho_\psi\,,
\end{equation}
where
\[
 \rho_\phi = 2q_1^2\alpha f_1^2\,,\quad\quad \rho_\psi = 2q_2(q_2\alpha-\omega)f_2^2\,.
\]
In order to show the "perfect" charge screening in this setting, integrate Eq.\ (\ref{eq:radeq3}) from zero to $\infty$, which yields, taking into account that $\alpha$ decays exponentially,
\begin{equation}\label{eq:screen}
0=\frac{4\pi}{e}\int_0^\infty\d r r^2 2\left[q_1^2\alpha f_1^2 + q_2(q_2\alpha-\omega)f_2^2\right]=Q_\phi + Q_\psi\,.
\end{equation}
As it has been analysed in the literature, Q-balls typically exist within a frequency interval
\begin{equation}\label{eq:frqr}
 \omega_{\rm min} < \omega < \omega_{\rm max}\,,
\end{equation}
where $\omega_{\rm min}$ and $\omega_{\rm max}$ are determined by the parameters of the theory.
In the present case these parameters are $\beta_1$, $\beta_2$, $\beta_{12}$, $\mu$, and the charges $q_i$.
We also note that one can set, for example, $\beta_{12}=1$ without losing generality.

The maximal frequency is determined by demanding $f_2\to0$ for large radii, to ensure the finiteness of the energy.
Since $f_2 \sim F_2\exp(-\sqrt{\mu-\omega^2} r)/r$ for $r\to\infty$ with
$F_2$ a constant, it follows that $\omega_{\rm max}=\sqrt{\mu}$.

The value of $\omega_{\rm min}$ is determined as follows \cite{ishiharaogawasoln, ishiharaogawasoln2}:
The effective potential, $U_{\rm eff}$ \eqref{eq:effpot} has critical points corresponding to the ``false'' vacuum $f_1=1$, $f_2=\alpha=0$ and to a ``true'' vacuum $f_1=f_1^0$, $f_2=f_2^0$, $\alpha=\alpha^0$. Near the minimal frequency, $\omega\approx\omega_{\rm min}$, the Q-ball tends to a homogenous ball filled by the ``true'' vacuum.
The minimal frequency is determined by the condition that the values of the effective potential are the same for the ``false'' resp.\ ``true'' vacua. For general values of the parameters of the theory, this leads to a somewhat complicated algebraic equation of degree 5 for $\omega_{\rm min}^2(\beta_{1},\beta_{2},\beta_{12},\mu,q_1,q_2)$.
In the special case when $\beta_2=0$ and $\mu=\beta_{12}$ one obtains an easily solvable cubic equation and finds
$\omega_{\rm min} = \sqrt{2}\sqrt{q\sqrt{2\beta_1\mu}-q^2\beta_1}$ ($q=q_2/q_1$).

Let us note another interesting special case considered in our paper: for $\beta_1=0$ it is found that
$\omega_{\rm min}=0$.

Yet another constraint for the existence of a Q-ball solution has been exhibited for the case $\beta_2=0$, $\mu=\beta_{12}$ in Refs.\ \cite{ishiharaogawasoln, ishiharaogawasoln2}, where it has been shown (in the case $q_1=q_2=1$) that
\begin{equation}\label{eq:alphapos}
 \beta_1 < \beta_{12}/2
\end{equation}
is necessary for the existence of a Q-ball solution, arising as a condition for the existence of the ``true'' vacuum with $(f_{1,2}^0)^2 \ge 0$.

By numerical evaluation of $\omega_{\rm min}(\beta_{1},\beta_{2},\beta_{12},\mu,q_1,q_2)$, we have found that for fixed values of $\beta_1$, $\beta_{12}$, and $\mu$, for values satisfying the criterion (\ref{eq:alphapos}), $\omega_{\rm min}$ is an increasing function of $\beta_2$. We have also found $\omega_{\rm min}$ to be an increasing function of $\mu$. This indicates that a non-zero $\beta_2$ decreases the domain of existence of the solutions (by increasing $\omega_{\rm min}$).

We also note that from a standard Derrick-type scaling argument ($r\rightarrow \lambda r$, Refs.\ \cite{derrick, rosen3}) one obtains from the effective action
$S_{\rm eff} = I_1 - I_3$ the virial relation
\begin{equation}\label{eq:virial}
I_1 = 3 I_3\,,
\end{equation}
where
\begin{equation}\label{eq:virial1}
I_1 =4\pi \int\d r  r^2 K_{\rm eff}\,,\quad I_3 = 4\pi \int\d r r^2 U_{\rm eff}\,.
\end{equation}

An interesting consequence of the virial relation \eqref{eq:virial1} is that
\begin{equation}\label{eq:virial-Energy}
\frac{E}{\eta}=-\omega \frac{Q_{\psi}}{q_2}+\frac{2}{3 e}I_1\,.
\end{equation}
The reason of Eq.\ (\ref{eq:virial-Energy}) only containing $Q_\psi$, and not both $Q_\phi$ and $Q_\psi$ in a symmetric way, is the gauge choice implicit in Ansatz \eqref{eq:Ansatz} the symmetry between $Q_\psi$ and $Q_\phi$ becomes manifest.
Using Ansatz \eqref{eq:Ansatz0} ,
\begin{equation}\label{eq:virial-Energy1}
\frac{E}{\eta}=-\left(\omega_1 \frac{Q_{\phi}}{q_1}+\omega_2 \frac{Q_{\psi}}{q_2}\right)+\frac{2}{3 e}I_1\,.
\end{equation}

\subsection{Numerical solutions}\label{ssec:num}
Q-ball solutions of the model defined by the action (\ref{eq:act}) conforming to the Ansatz (\ref{eq:Ansatz}) have been obtained by the numerical solution of the radial equations (\ref{eq:radeq1}-\ref{eq:radeq3}), using the {\sc Colnew} package \cite{colnew, ascher} implementing collocation on Gaussian points. We have used an interval $0 \le r \le r_{\rm max}$, which was large enough for the radial functions to reach their limiting values within numerical precision.

\begin{figure}[h!]
\begin{subfigure}[t]{0.5\textwidth}
 \noindent\hfil\includegraphics[scale=.5]{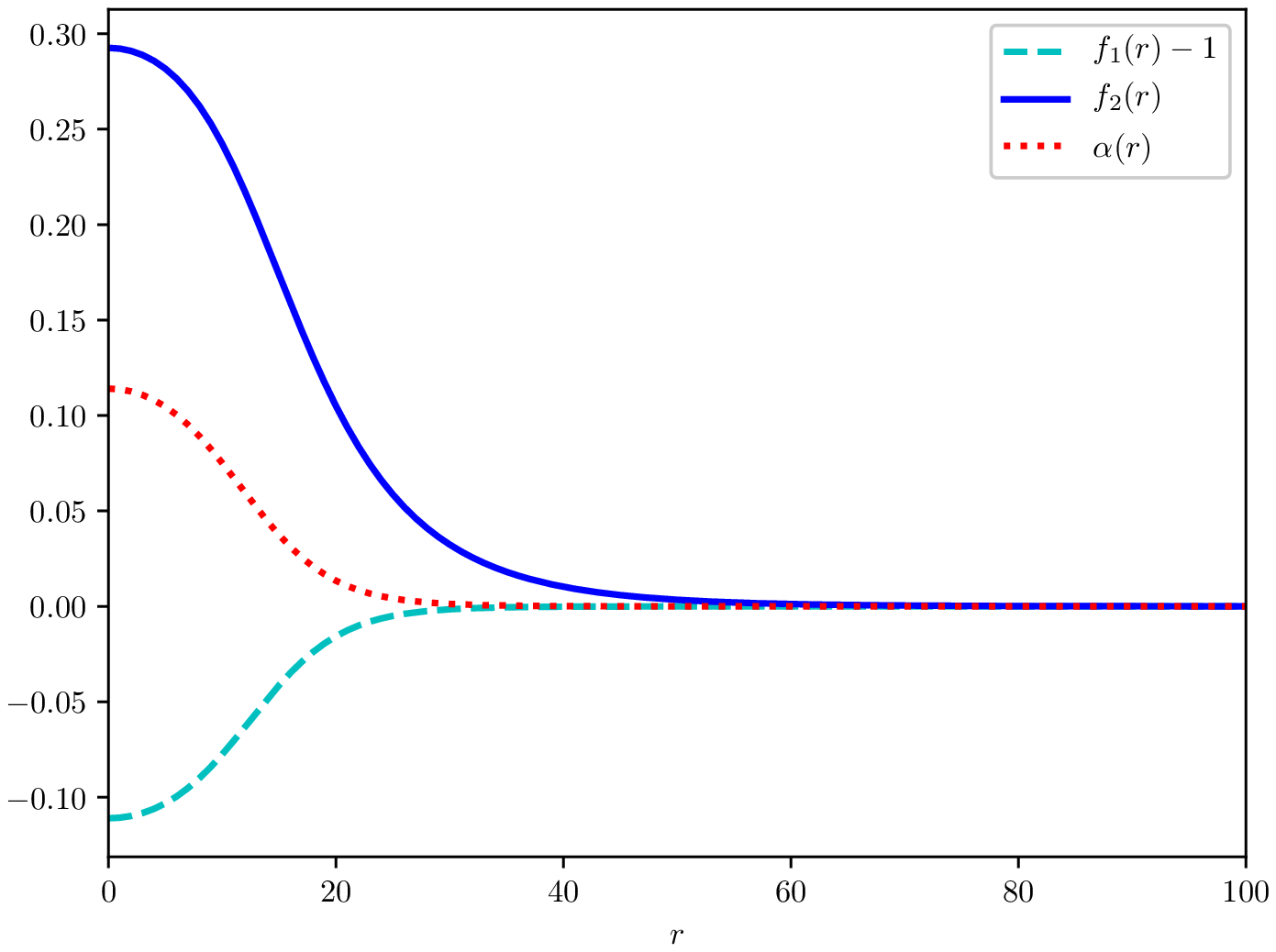}
 \caption{}
 \label{fig:beta2xmplef}
\end{subfigure}
\begin{subfigure}[t]{0.5\textwidth}
 \noindent\hfil\includegraphics[scale=.5]{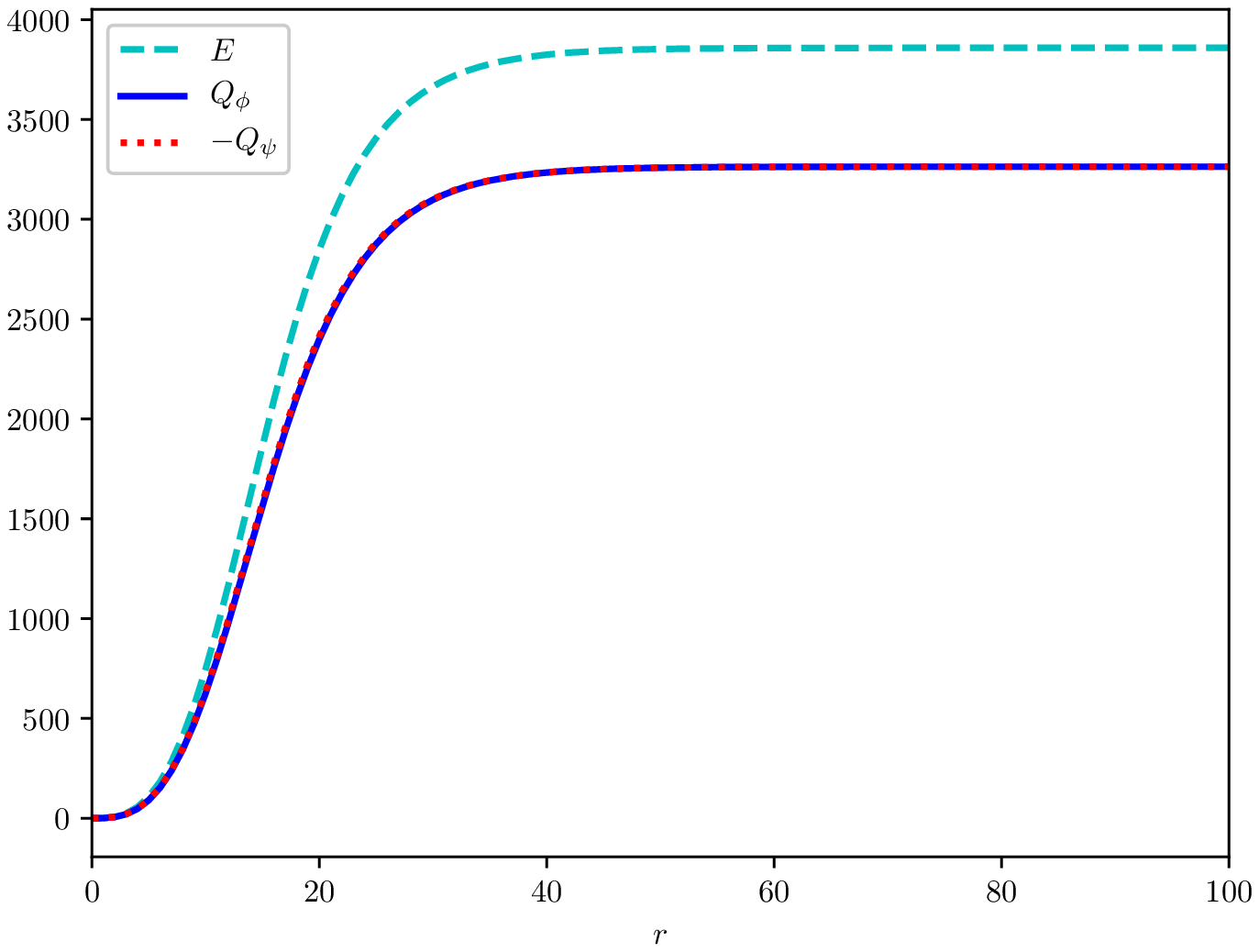}
 \caption{}
\label{fig:beta2xmplee}
\end{subfigure}

\caption{(a) The profile functions of a typical Q-ball; $\beta_1=0.5$, $\beta_{12}=\mu=1.4$, $\beta_2=0.25$, $\omega=1.180$. (b) The energy and charge distributions of the same Q-ball.}
\label{fig:beta2xmple}
\end{figure}

In Fig.\ \ref{fig:beta2xmplef}, an example of a Q-ball in the full non-linear model is shown. The visual appearance of solutions with and without the quartic self-interaction term of the field $\psi$ are almost identical. In Fig.\ \ref{fig:beta2xmplee}, the energy and charge distributions of the solution of the non-linear model are displayed. The cancellation of the charge densities due of the two fields is again local.

\begin{figure}[h!]
\begin{subfigure}[t]{0.5\textwidth}
 \noindent\hfil\includegraphics[scale=.5]{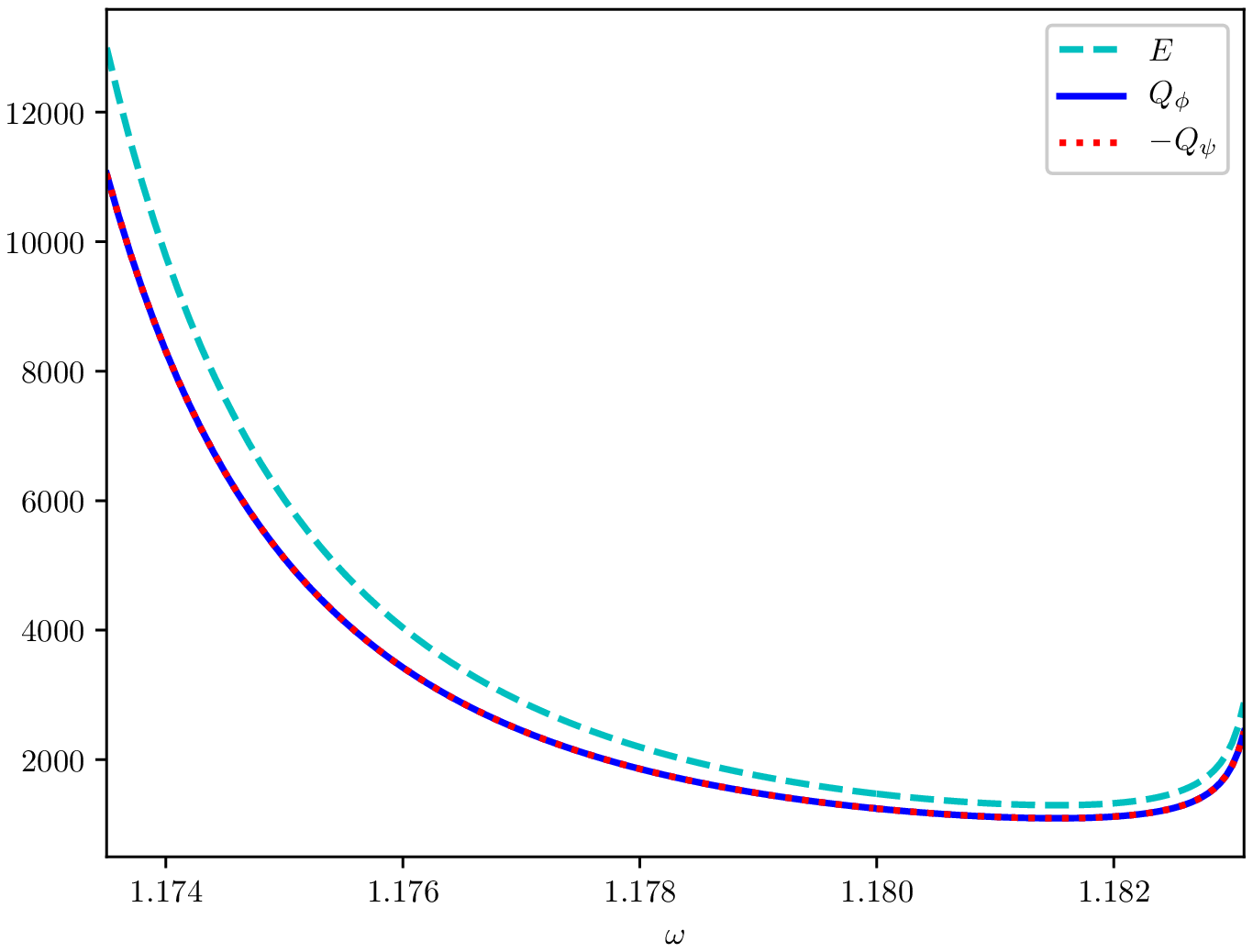}
 \caption{$\beta_2=0$}
\end{subfigure}
\begin{subfigure}[t]{0.5\textwidth}
 \noindent\hfil\includegraphics[scale=.5]{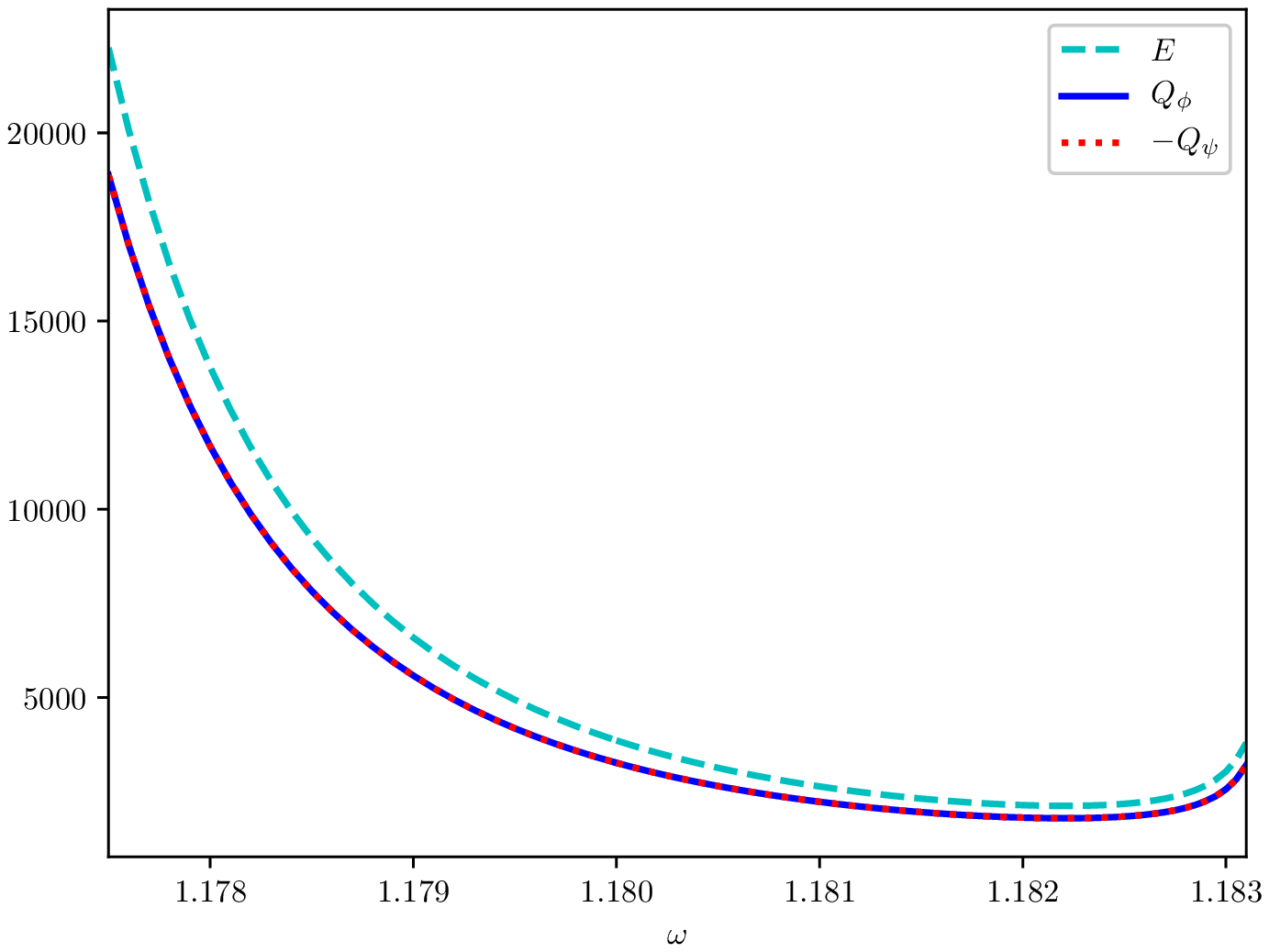}
 \caption{$\beta_2=0.25$}
\end{subfigure}
\caption{The frequency dependence of the conserved quantities, $\beta_1=0.5$, $\beta_{12}=\mu=1.4$, (a) $\beta_2=0$ and (b) $\beta_2=0.25$.}
\label{fig:omswp}
\end{figure}

\begin{figure}[h!]
\begin{subfigure}[t]{0.5\textwidth}
 \noindent\hfil\includegraphics[scale=.5]{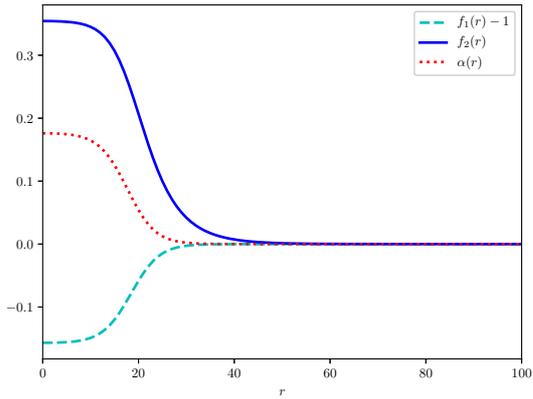}
 \caption{$\omega=1.174$}
\end{subfigure}
 \begin{subfigure}[t]{0.5\textwidth}
 \noindent\hfil\includegraphics[scale=.5]{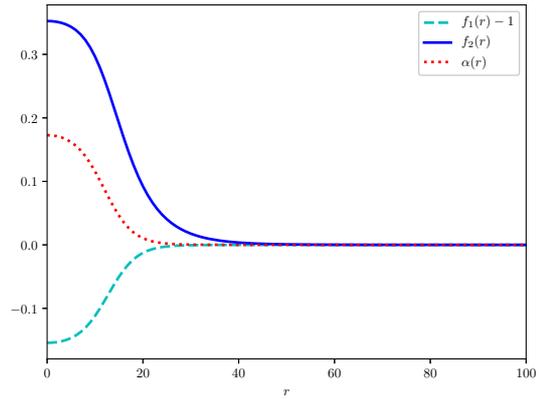}
 \caption{$\omega=1.176$}
\end{subfigure}
\begin{subfigure}[t]{0.5\textwidth}
 \noindent\hfil\includegraphics[scale=.5]{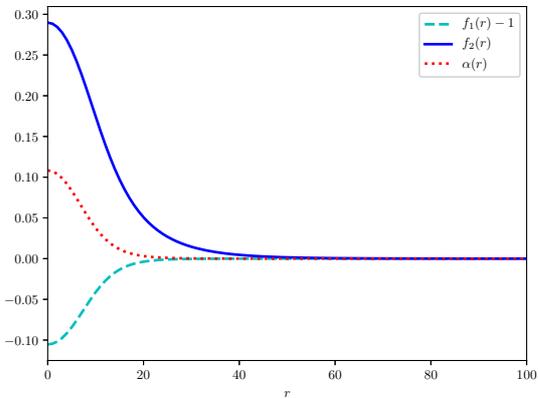}
 \caption{$\omega=1.18$}
\end{subfigure}
\begin{subfigure}[t]{0.5\textwidth}
 \noindent\hfil\includegraphics[scale=.5]{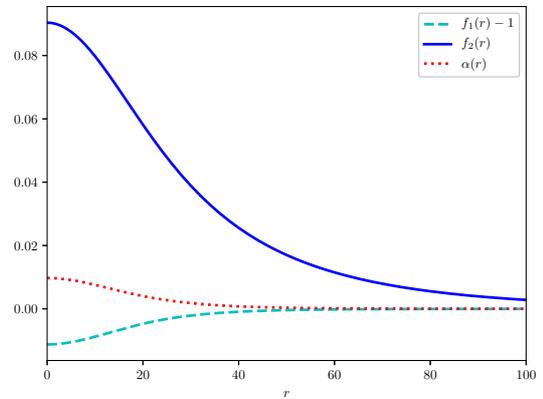}
 \caption{$\omega=1.183$}
\end{subfigure}
\caption{The frequency dependence of the solutions approaching the frequency limits, $\omega_{\rm min}$ and $\omega_{\rm max}$, $\beta_1=0.5$, $\beta_{12}=\mu=1.4$, $\beta_2=0$.}
\label{fig:4om}
\end{figure}

We have chosen the parameters $\beta_1=0.5$, $\beta_{12}=\mu=1.4$ as in Refs.\ \cite{ishiharaogawasoln, ishiharaogawasoln2}\footnote{To translate from the conventions of Refs.\ \cite{ishiharaogawasoln, ishiharaogawasoln2} to ours, the replacement $\lambda\to2\lambda_1$ shall be performed, i.e., $\lambda=1$ in Refs.\ \cite{ishiharaogawasoln, ishiharaogawasoln2} corresponds to $\lambda=1/2$ here.}, and considered the numerically available frequency range $1.174 \le \omega \le 1.183$. For each value of the frequency $\omega$, we calculated the Q-ball solutions for a range of the parameter $\beta_2$, starting from $\beta_2=0$, increasing $\beta_2$ to upper values of the order 0.1. In these ranges, we have found that the energy $E$ and the magnitude of the charges $|Q_\phi|$ and $|Q_\psi|$ were monotonically increasing functions of the parameter $\beta_2$. The frequency dependence of the energy and charges is depicted in Fig.\ \ref{fig:omswp}. In Figs.\ \ref{fig:4om}(a)-(c) we present the change of solutions when approaching the lower limiting frequency (see also Refs. \cite{ishiharaogawasoln, ishiharaogawasoln2}). Upon approaching the minimal frequency, the difference between the effective potential $U_{\rm eff}$ of the false vacuum at the origin and the real one becomes smaller, and, therefore, due to a Derrick-type argument, the gradient terms must become small as well, and the Q-ball expands.

Approaching the upper frequency limit, $\omega_{\rm max} = \sqrt\mu$, the exponent of the asymptotic radial decay of $f_2$ becomes small, and the asymptotic tail of the solutions in $f_2$ expands, in contrast to the lower limit, where the core of the solution expands, see Figs.\ \ref{fig:4om}(c) and (d).

\begin{figure}[h!]
\begin{subfigure}[t]{0.5\textwidth}
 \noindent\hfil\includegraphics[scale=.5]{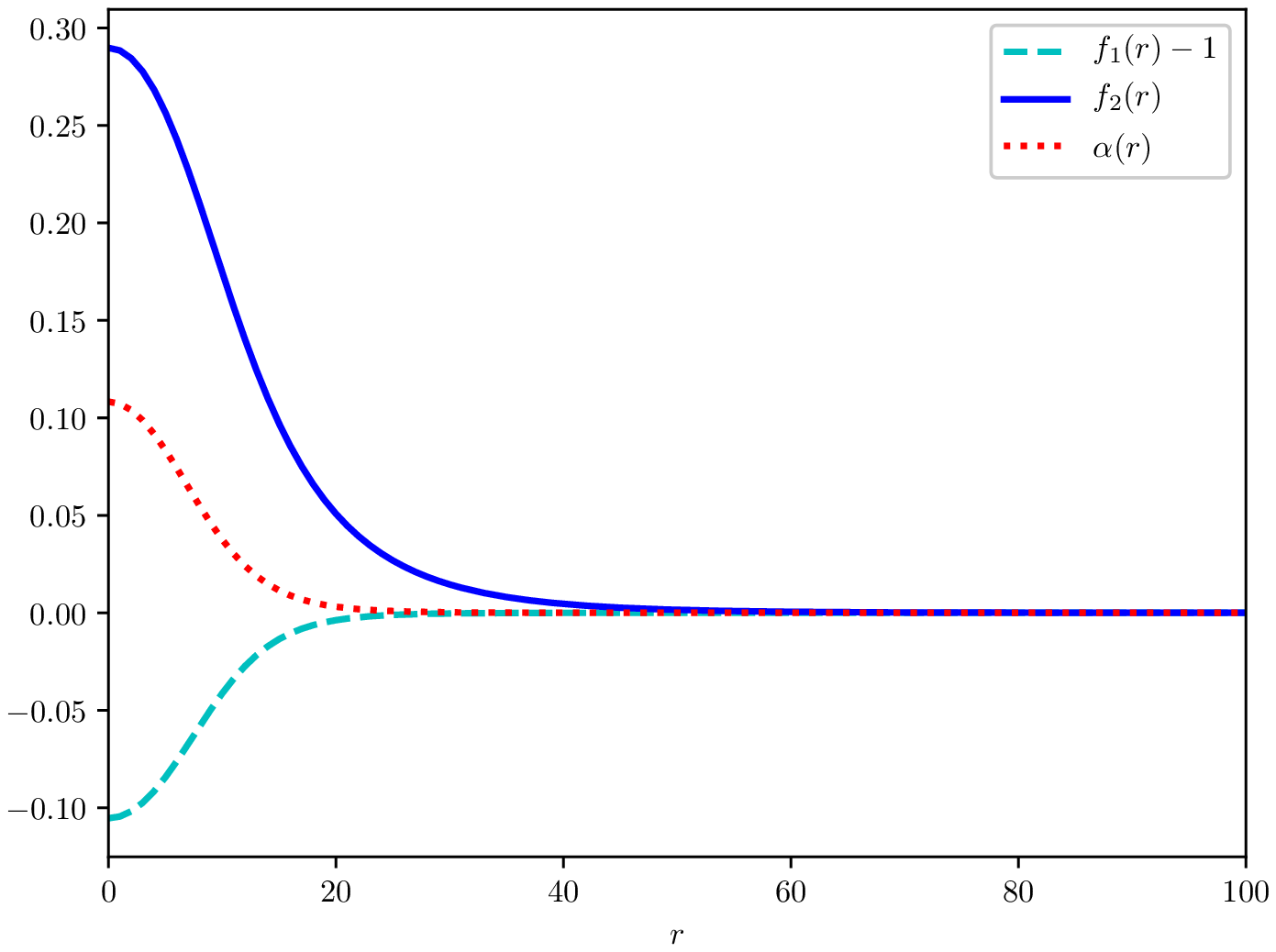}
 \caption{$\beta_2=0$}
\end{subfigure}
\begin{subfigure}[t]{0.5\textwidth}
 \noindent\hfil\includegraphics[scale=.5]{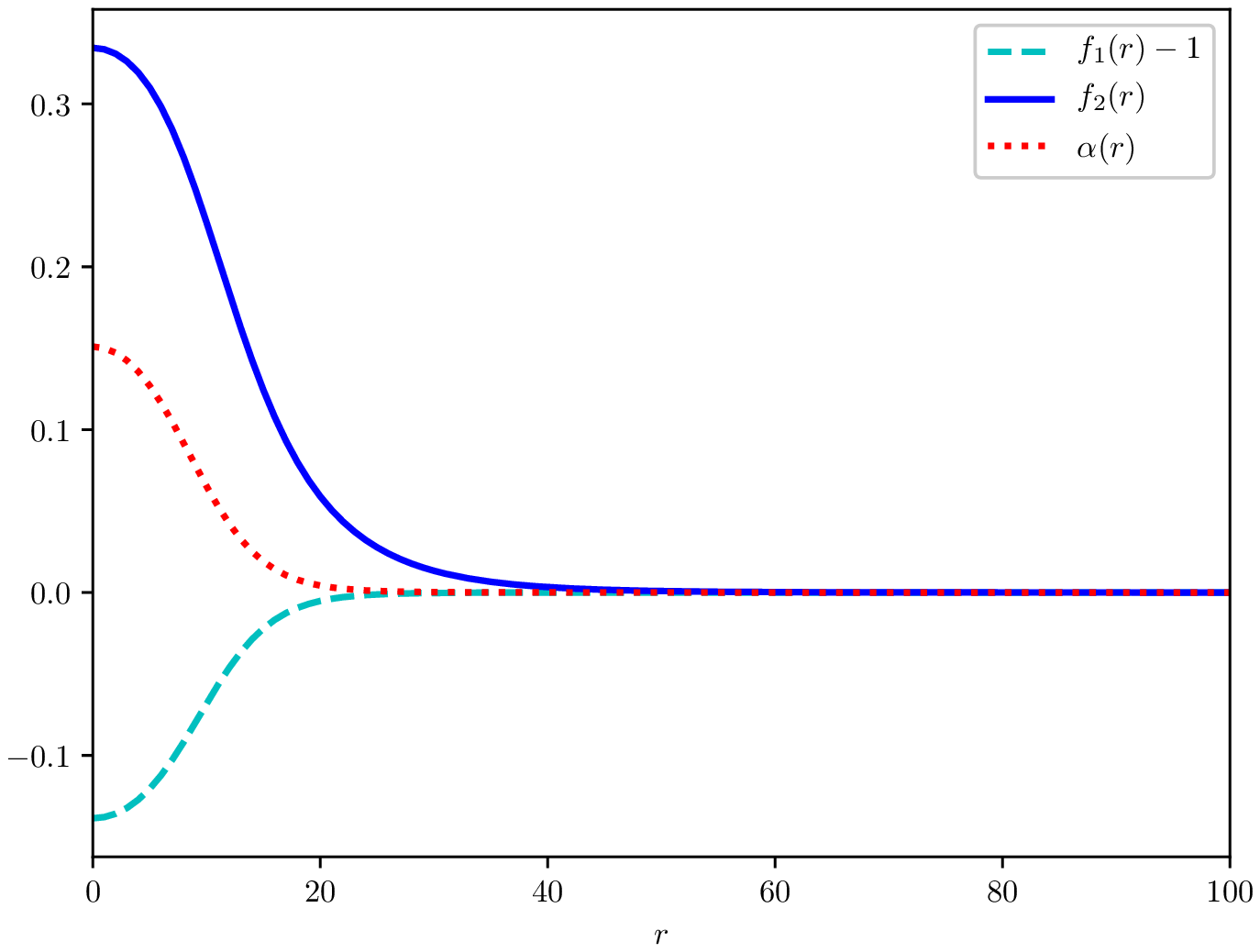}
 \caption{$\beta_2=0.15$}
\end{subfigure}
 \begin{subfigure}[t]{0.5\textwidth}
 \noindent\hfil\includegraphics[scale=.5]{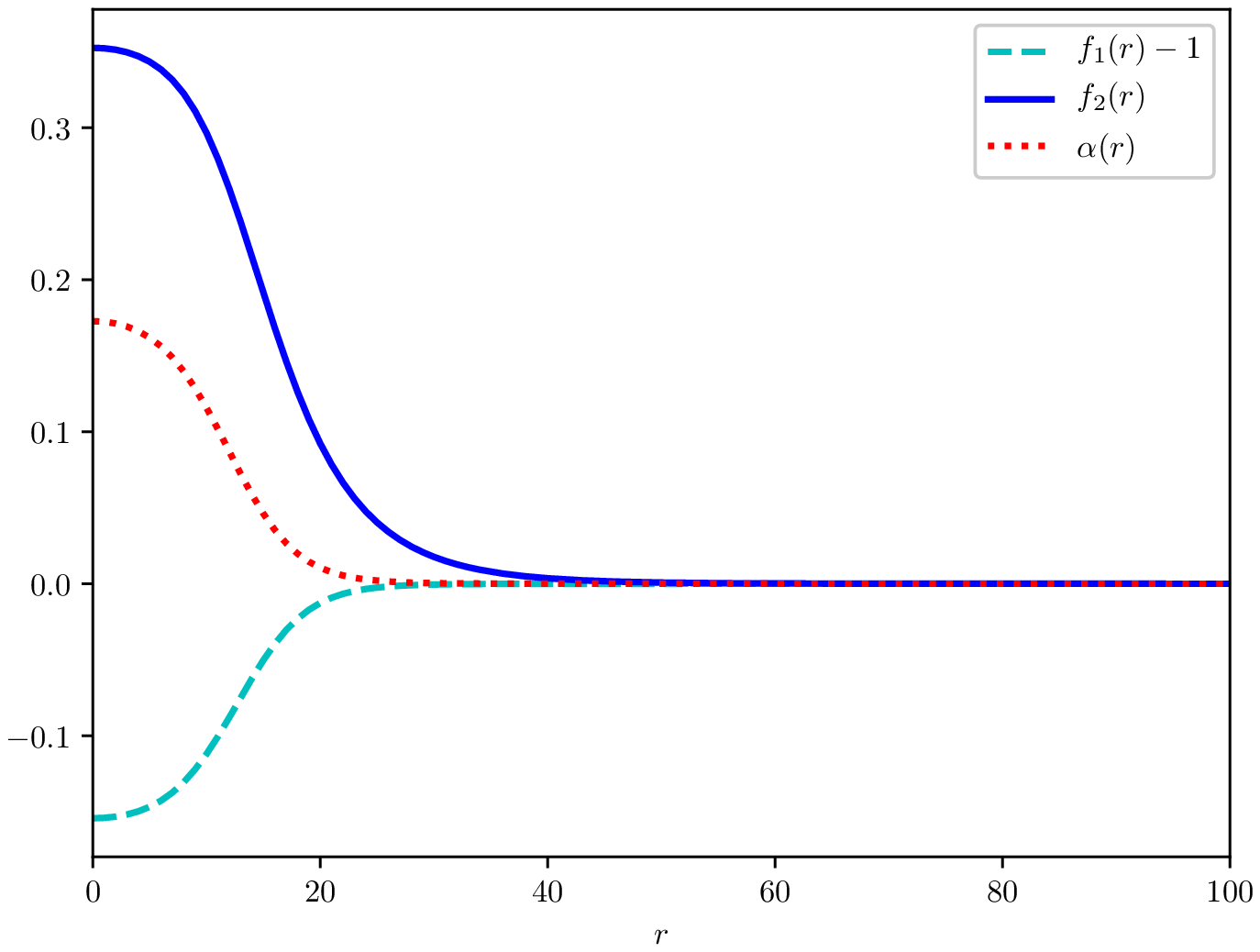}
 \caption{$\beta_2=0.3$}
\end{subfigure}
\begin{subfigure}[t]{0.5\textwidth}
 \noindent\hfil\includegraphics[scale=.5]{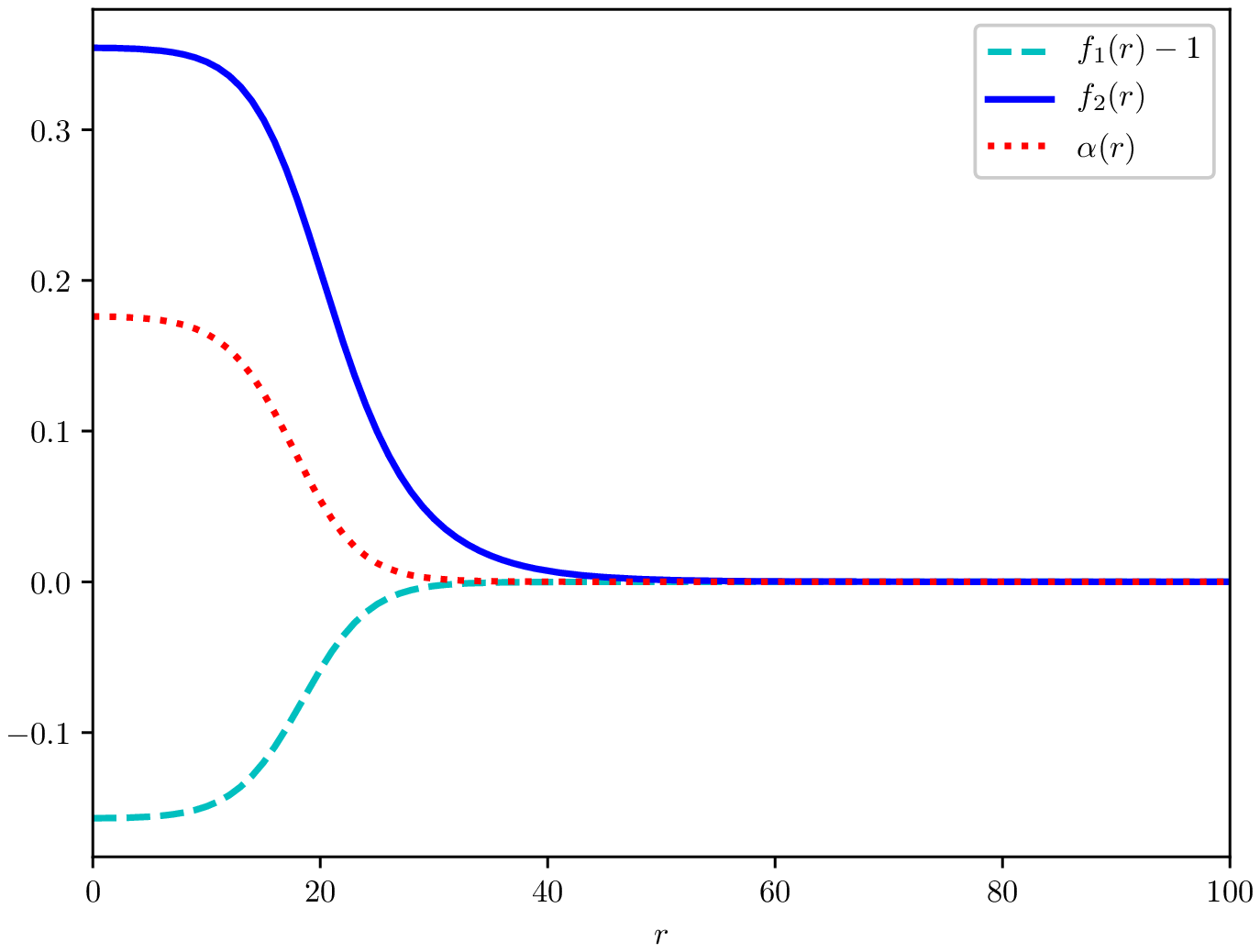}
 \caption{$\beta_2=0.45$}
\end{subfigure}
\caption{The dependence of the solutions on the parameter $\beta_2$, $\beta_1=0.5$, $\beta_{12}=\mu=1.4$, $\omega=1.18$.}
\label{fig:4beta2}
\end{figure}

\begin{figure}[h!]
\begin{subfigure}[t]{0.5\textwidth}
 \noindent\hfil\includegraphics[scale=.5]{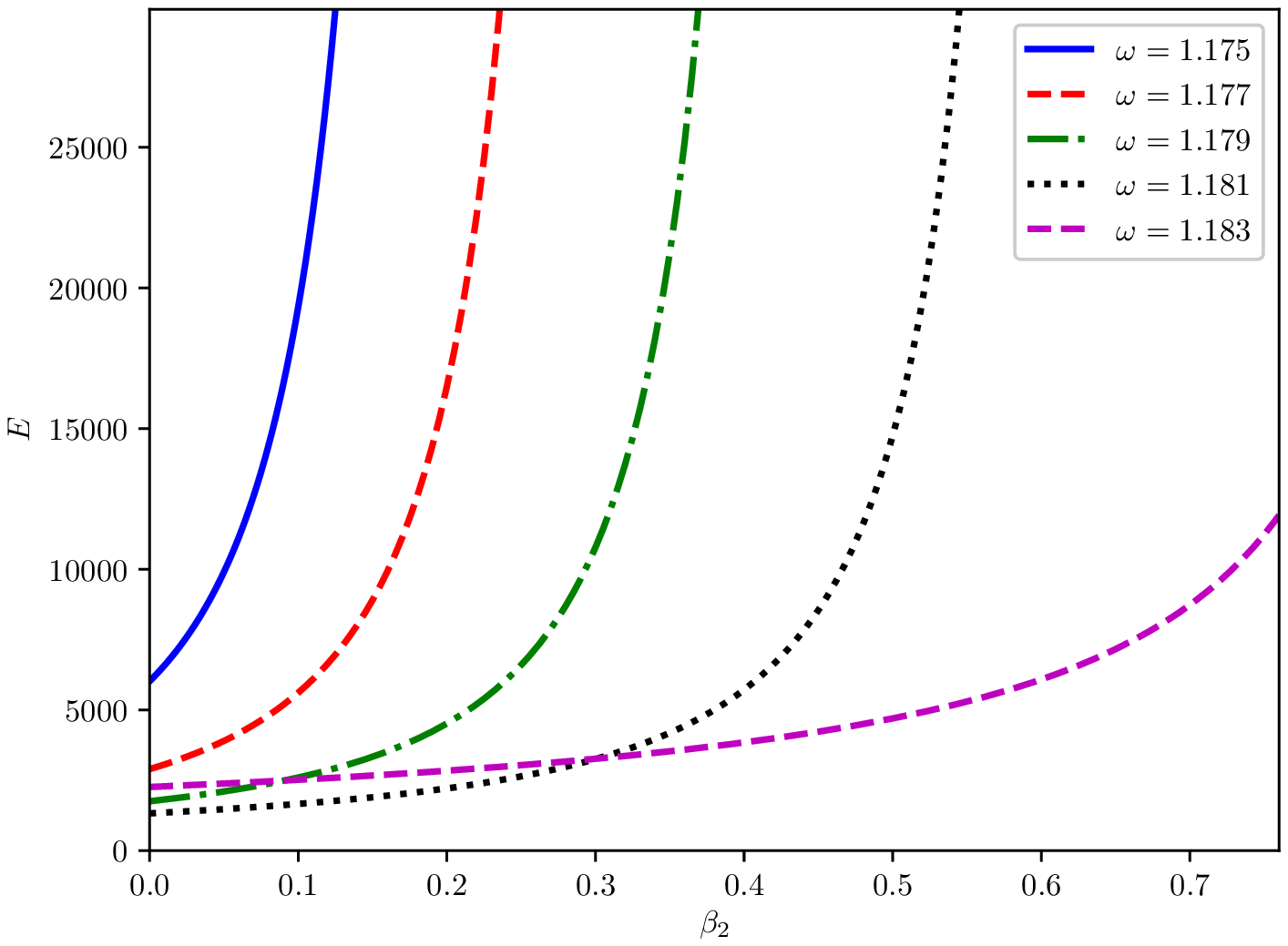}
 \caption{}
\end{subfigure}
\begin{subfigure}[t]{0.5\textwidth}
 \noindent\hfil\includegraphics[scale=.5]{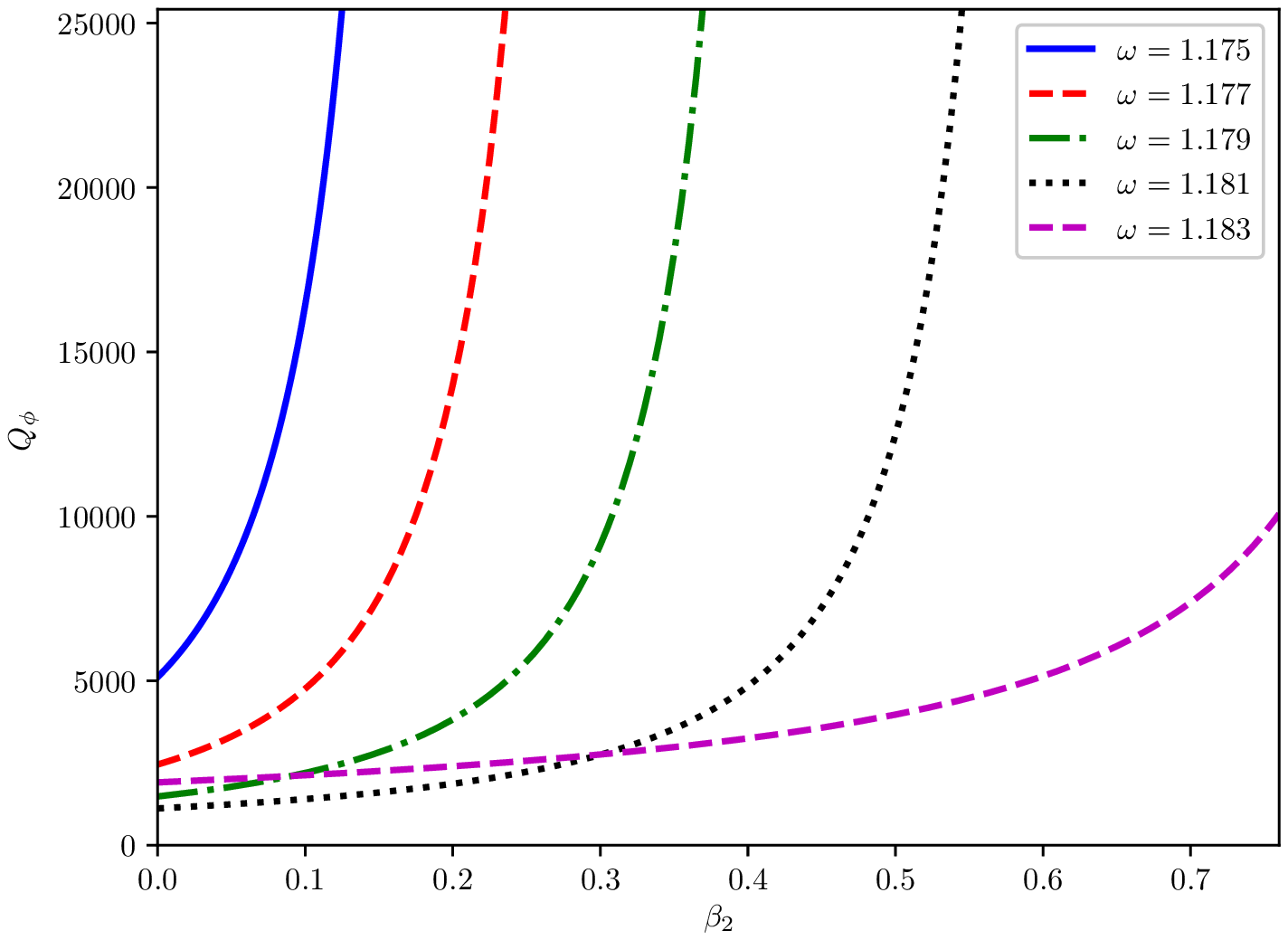}
 \caption{}
\end{subfigure}
\caption{The $\beta_2$ parameter dependence of (a) the energy $E$ and (b) the charge $Q_\phi$, $\beta_1=0.5$, $\beta_{12}=\mu=1.4$.}
\label{fig:b2swp}
\end{figure}

The behaviour for large $\beta_2$ is similar to the one for the frequency approaching the minimal one; see Fig.\ \ref{fig:4beta2}. The core of the Q-ball expands, and the energy and the charge diverge. The energy $E$ and the charge $Q_\phi$ are depicted as a function of $\beta_2$ in Fig.\ \ref{fig:b2swp}. Stability of the theory requires $\beta_2 \ge 0$, and for $\beta_2 =0$, the solutions of Refs.\ \cite{ishiharaogawasoln, ishiharaogawasoln2} are recovered.

\begin{figure}
\begin{subfigure}[t]{0.5\textwidth}
 \noindent\hfil\includegraphics[scale=.5]{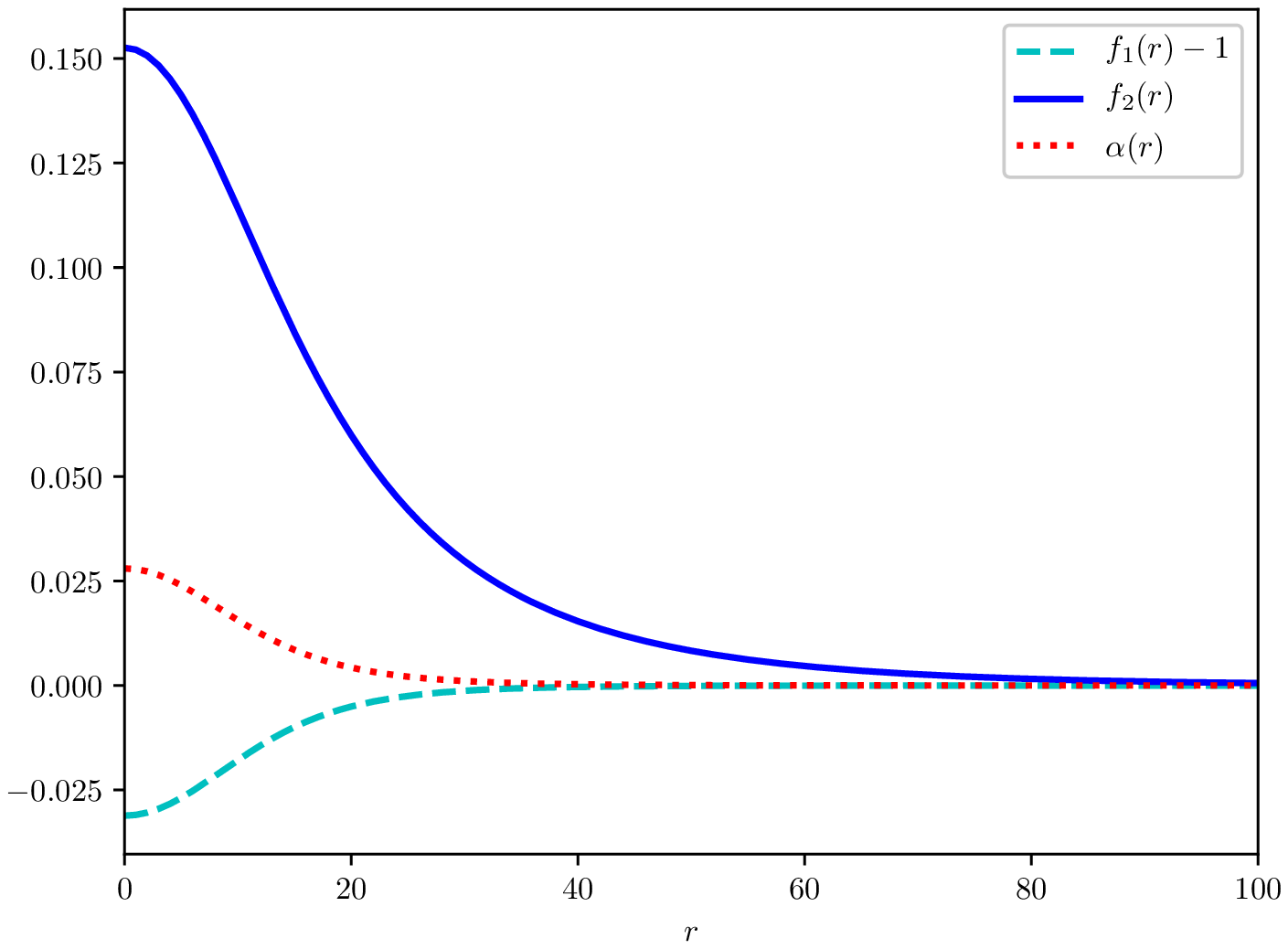}
 \caption{$\mu=1.413$}
 \label{fig:mud}
\end{subfigure}
\begin{subfigure}[t]{0.5\textwidth}
 \noindent\hfil\includegraphics[scale=.5]{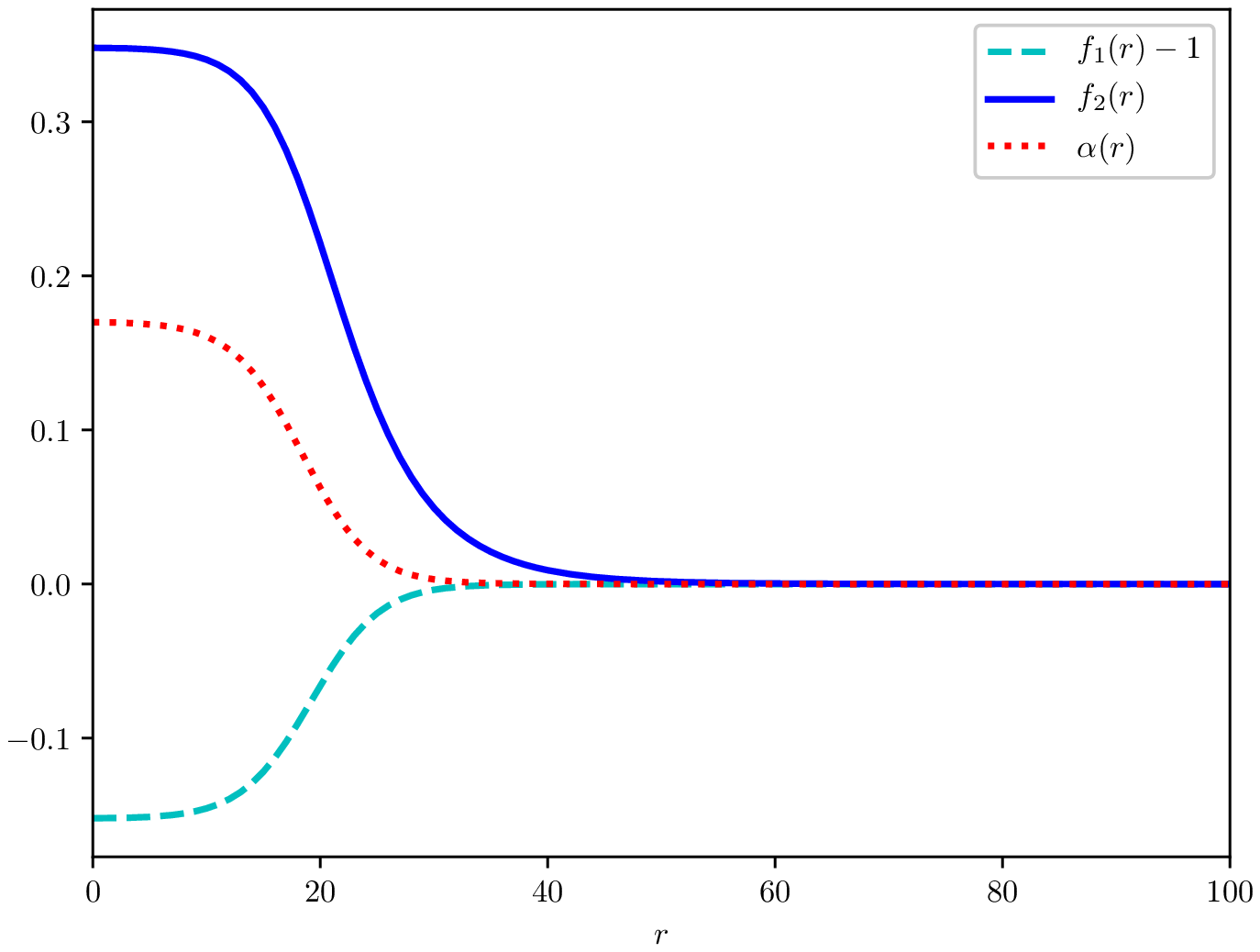}
 \caption{$\mu=1.396$}
 \label{fig:muu}
\end{subfigure}
\caption{The dependence of the solutions on the parameter $\mu$ upon approaching  (a) the maximal value, determined by $\omega_{\rm min}$ reaching $\omega$ and (b) the minimal value determined by $\omega^2$.}
\label{fig:mulimits}
\end{figure}

\begin{figure}[h!]
\begin{subfigure}[t]{0.5\textwidth}
 \noindent\hfil\includegraphics[scale=.5]{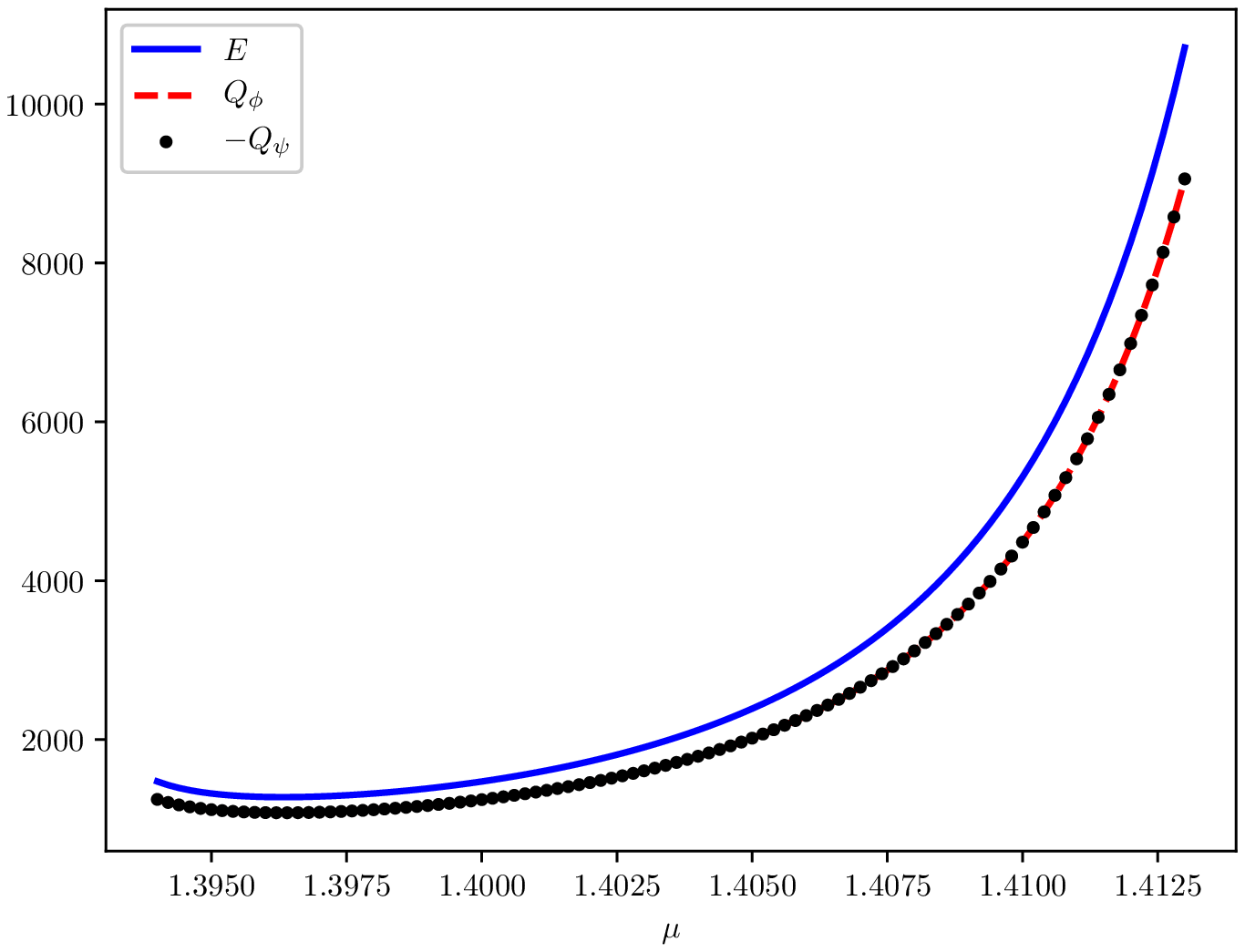}
 \caption{$\beta_2=0$}
\end{subfigure}
\begin{subfigure}[t]{0.5\textwidth}
 \noindent\hfil\includegraphics[scale=.5]{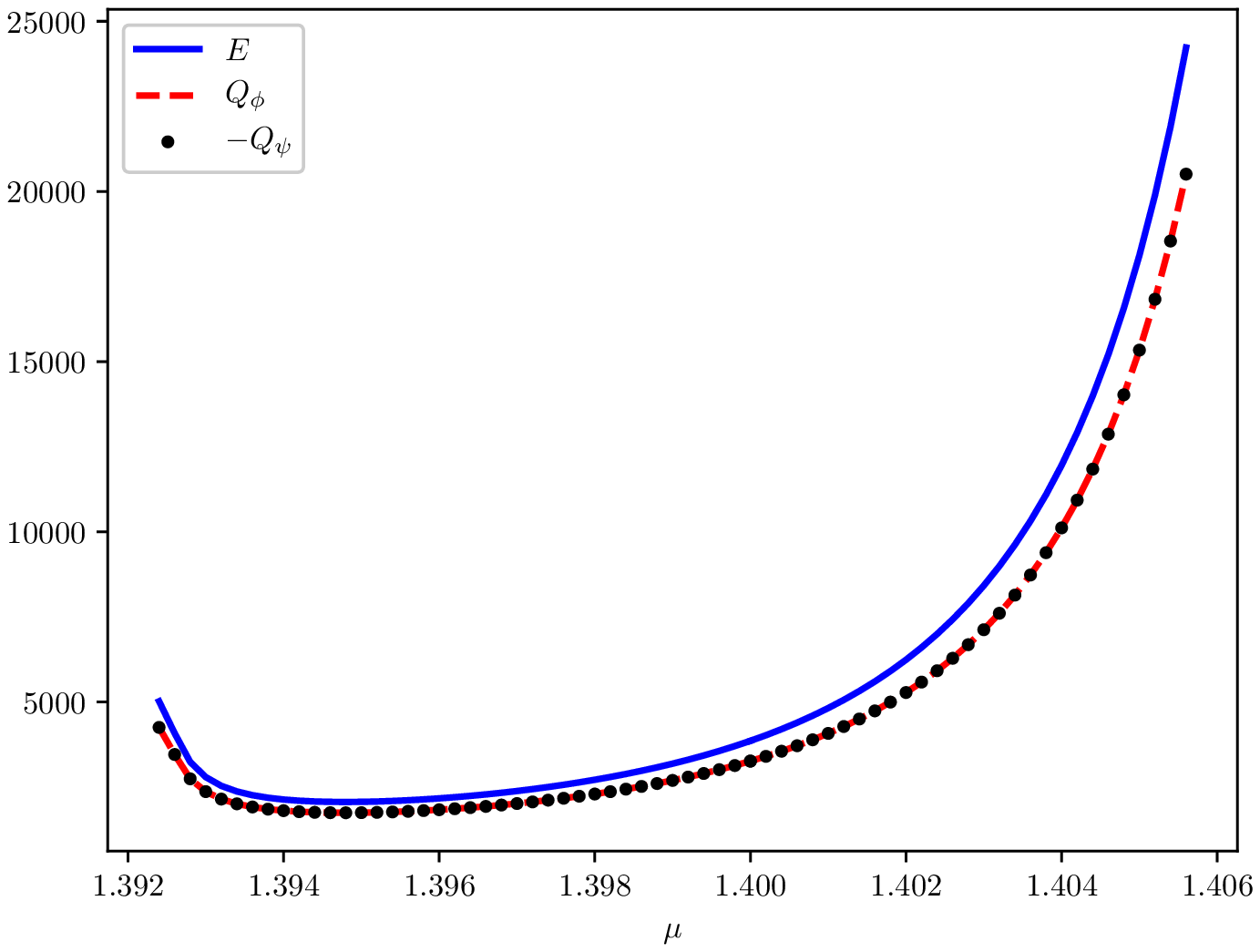}
 \caption{$\beta_2=0.25$}
\end{subfigure}
\caption{The dependence of the conserved quantities on the parameter $\mu$, $\beta_1=0.5$, $\beta_{12}=\mu=1.4$, $\omega=1.18$, (a) $\beta_2=0$ and (b) $\beta_2=0.25$.}
\label{fig:muswp}
\end{figure}

Varying the parameter $\mu$ is quite similar to the above ones. There is a lower limit due to the exponent of the radial decay of the profile function $f_2$, $\mu_{\rm min} = \omega^2$, and upon approaching it, the tail of the solution in $f_2$ expands, and not the core; see Fig.\ \ref{fig:mud}. There is also an upper limit, $\mu=\mu_{\rm max}$, where the core expands, depicted in Fig.\ \ref{fig:muu}. The dependence of the integrated quantities $E$, $Q_\phi$ and $Q_\psi$ is depicted in Fig.\ \ref{fig:muswp}

\begin{figure}[h!]
\begin{subfigure}[t]{0.5\textwidth}
 \noindent\hfil\includegraphics[scale=.5]{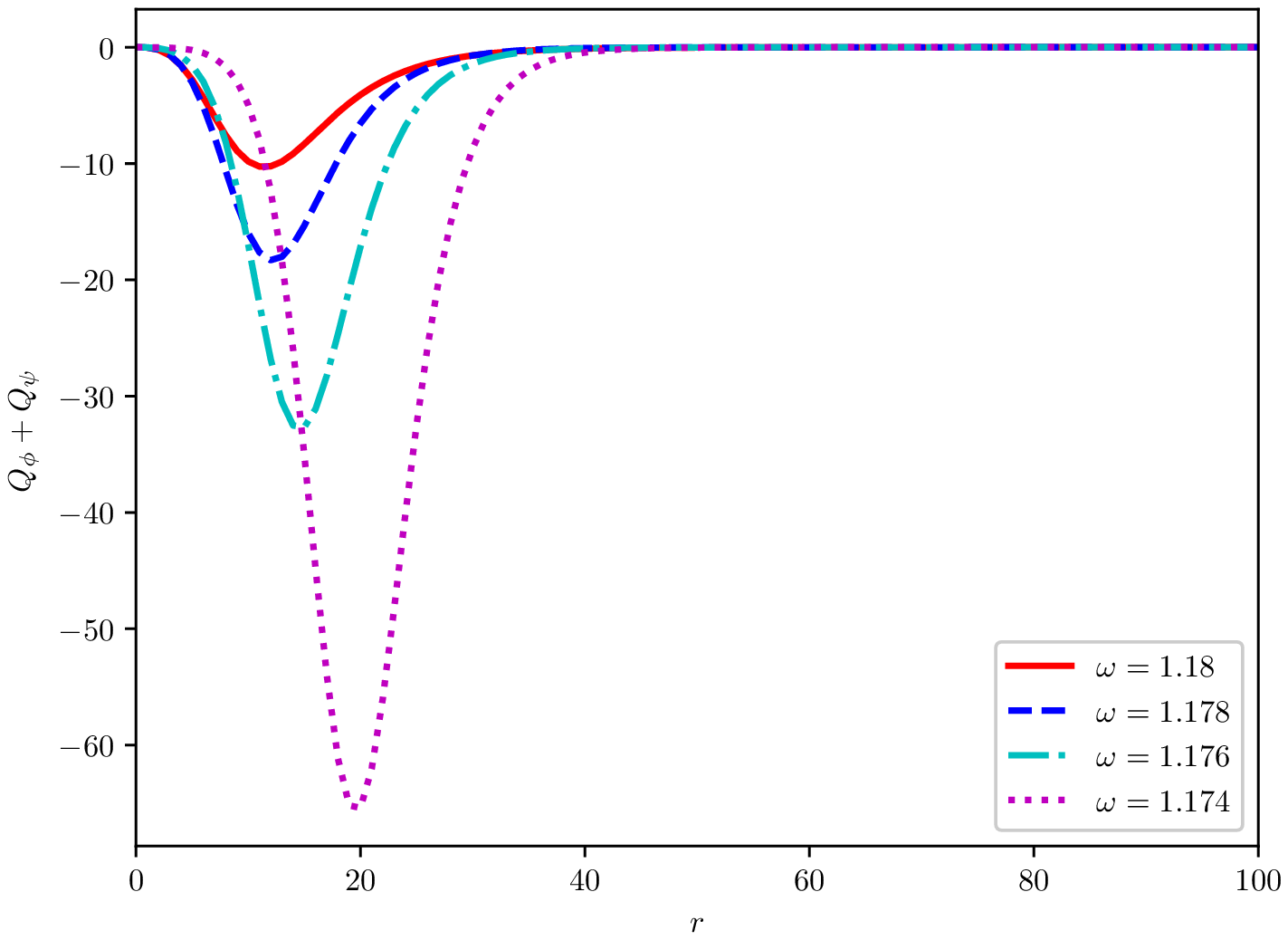}
 \caption{Towards $\omega_{\rm min}$}
\end{subfigure}
\begin{subfigure}[t]{0.5\textwidth}
 \noindent\hfil\includegraphics[scale=.5]{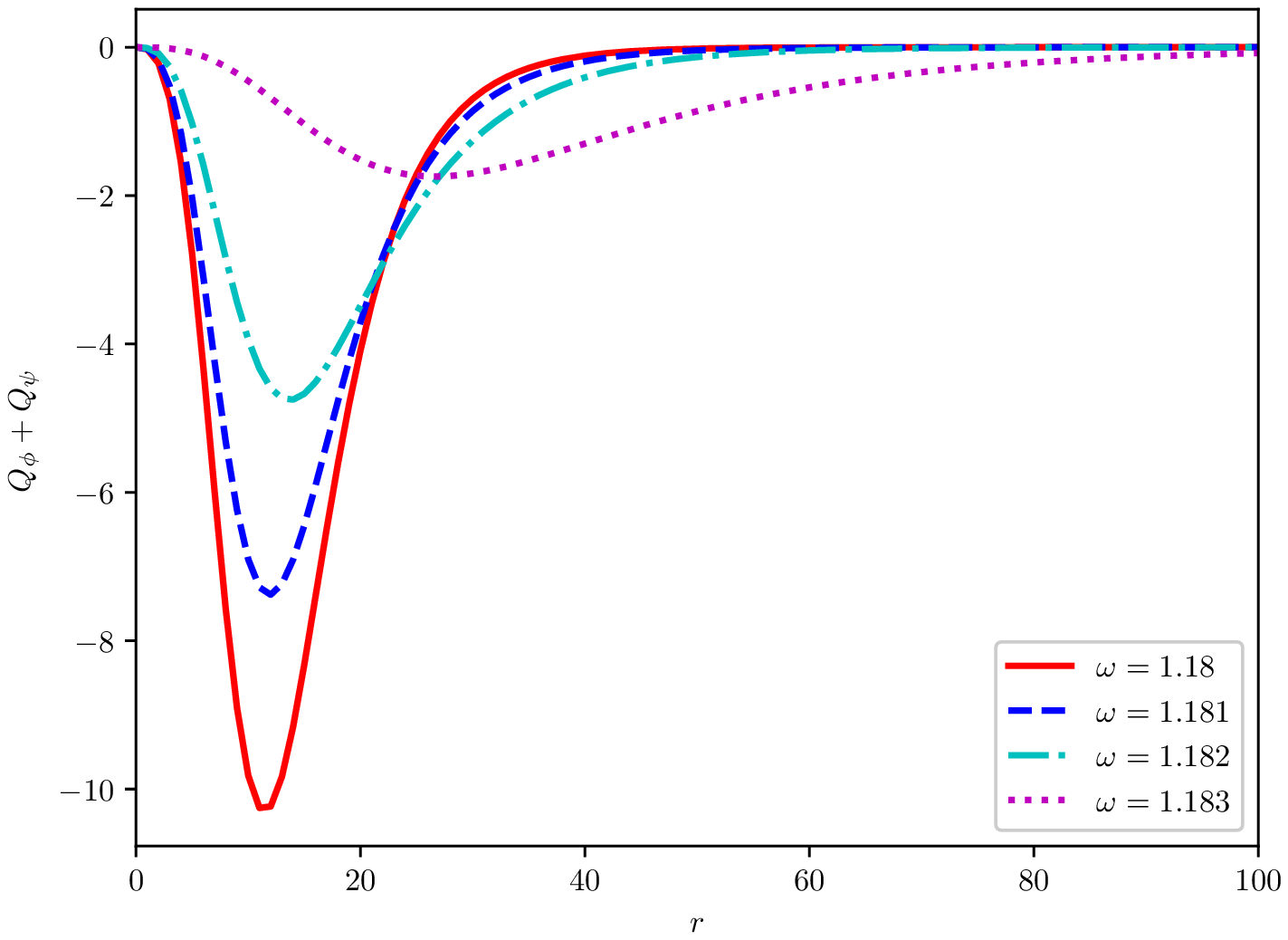}
 \caption{Towards $\omega_{\rm max}$}
\end{subfigure}
\caption{The dependence of the screened (summed over species) charge $Q_\phi+Q_\psi$ on the frequency $\omega$, $\beta_1=0.5$, $\beta_{12}=\mu=1.4$, $\beta_2=0$, upon (a) decreasing $\omega$ and (b) increasing $\omega$.}
\label{fig:chgsum}
\end{figure}

In agreement with our analytical result \eqref{eq:screen} we have found, that the total charges $Q_\phi$ and $Q_\psi$ cancel each other for all solutions considered within numerical precision.
In addition, we have considered the precision of the cancellation of the charges as a function of the radius, for various parameters. In Fig.\ \ref{fig:chgsum} we have depicted the charge remaining after the cancellation for different frequencies. It can be seen, that the precision of the (local) cancellation decreases with lower frequencies, and the remaining charge is sharply peaked, and the peak moves to larger radii when the frequency $\omega$ approaches the lower limit, $\omega_{\rm min}$. Upon approaching the upper limit $\omega_{\rm max}=\sqrt{\mu}$, the remaining charge becomes smaller, but remains there for larger radii, in a less peaked shape.

The dependence on other parameters can be explained by considering what happens when changing the given parameter. When the parameter increases $\omega_{\rm min}$, the same happens (for fixed frequency) as when $\omega$ approaches $\omega_{\rm min}$; this is the case for increased $\mu$ and $\beta_2$. On the contrary, upon lowering $\mu$, the solutions behave as if $\omega$ was increased.

\subsection{Stability of the solutions}\label{ssec:soln}
A detailed analysis of the stability of Q-balls is still a challenging subject, with some open questions (see the Review \cite{NSrev}). However, an approximate indication of stability can be obtained by comparing the energy $E$ of the solution to the energy $E_{\rm free}$ of free $\psi$ particles with the same charge \cite{LeePang}; if the energy of free $\psi$ particles is larger, it is energetically not favourable for the soliton to fall apart into free particles,
where
\begin{equation}\label{eq:Efree}
 N = Q_\psi/q_2\,,\quad E_{\rm free} = \sqrt{\mu}N\,.
\end{equation}
In addition, if the ratio $E/E_{\rm free}$ is not only below 1, but a decreasing function of $N$, it is also not favourable for the Q-ball to split into smaller Q-balls.

\begin{figure}[h!]
\begin{subfigure}[t]{0.5\textwidth}
 \noindent\hfil\includegraphics[scale=.5]{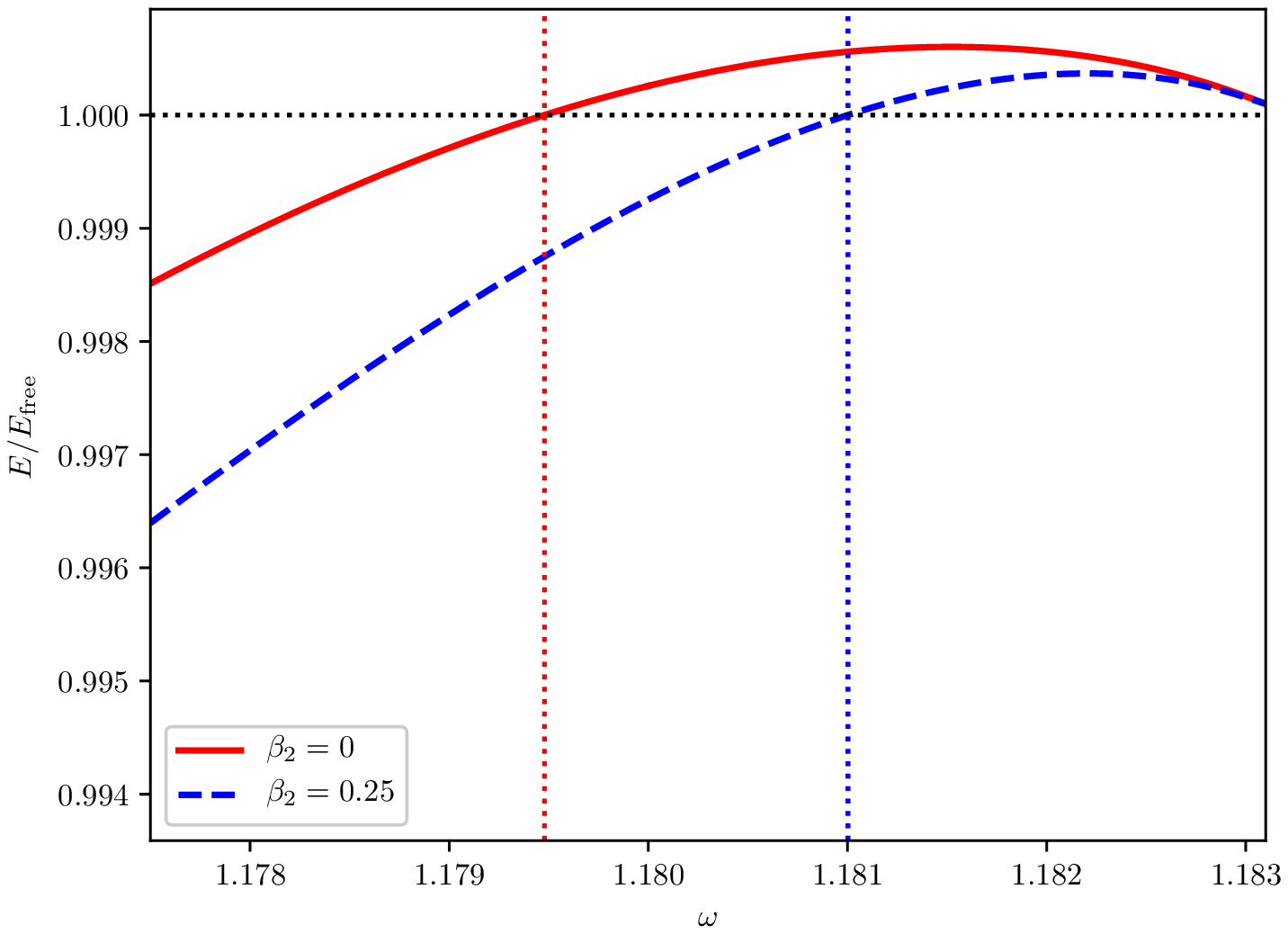}
 \caption{}
 \label{fig:stabEEom}
\end{subfigure}
\begin{subfigure}[t]{0.5\textwidth}
 \noindent\hfil\includegraphics[scale=.5]{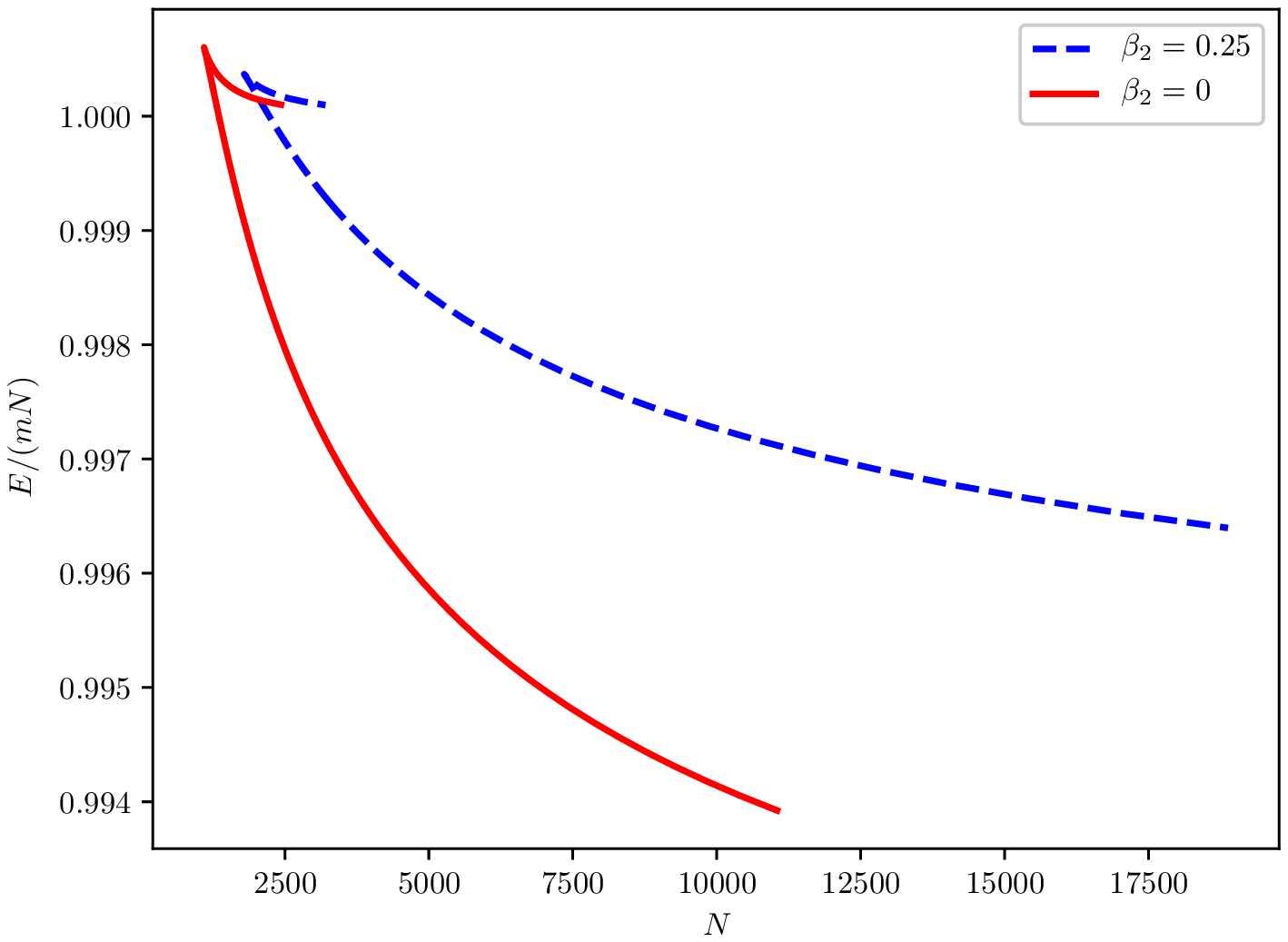}
 \caption{}
 \label{fig:stabEEN}
\end{subfigure}
\caption{The ratio of the energy of the solutions and that of free $\psi$ particles of the same charge, for $\beta_1=0.5$, $\beta_2=0.25$ and 0, $\beta_{12}=\mu=1.4$, (a) as a function of the frequency $\omega$ and (b) of the number of $\psi$ particles in the Q-ball.}
\label{fig:stab}
\end{figure}

The ratio $E/E_{\rm free}$ is depicted in Fig.\ \ref{fig:stabEEom} as a function of $\omega$. The behaviour for nonzero quartic coupling $\beta_2$ is the same as found in Refs.\ \cite{ishiharaogawasoln, ishiharaogawasoln2} for $\beta_2=0$, i.e., below a critical frequency $\omega=\omega_{\rm cr}$, $E/E_{\rm free} <1$, indicating that the solutions are stable against falling apart into free particles (with arbitrarily large energy and charge, when approaching $\omega_{\rm min}$).  At $\beta_1 =0.5$, $\beta_{12}=\mu=1.4$, we have found that for $\beta_2=0.25$, $\omega_{\rm cr} = 1.1810$ and for $\beta_2=0$, $\omega_{\rm cr} = 1.1795$.

Fig.\ \ref{fig:stabEEN} depicts the energy ratio as a function of the number of $\psi$ particles. Note, that on the branch of the figure corresponding to $\omega<\omega_{\rm cr}$, the function is monotonically decreasing, indicating that the corresponding Q-balls are also stable against falling apart into smaller ones.

\subsection{The effect of varying the charges}\label{ssec:e1e2}

We have also considered the effect of varying the charges $q_1$ resp.\  $q_2$ of the fields $\phi$ and $\psi$. In this section we restrict ourselves to the case $\beta_2=0$ and $\beta_{12}=\mu$.

\begin{figure}[h!]
\begin{subfigure}[t]{0.5\textwidth}
 \noindent\hfil\includegraphics[scale=.5]{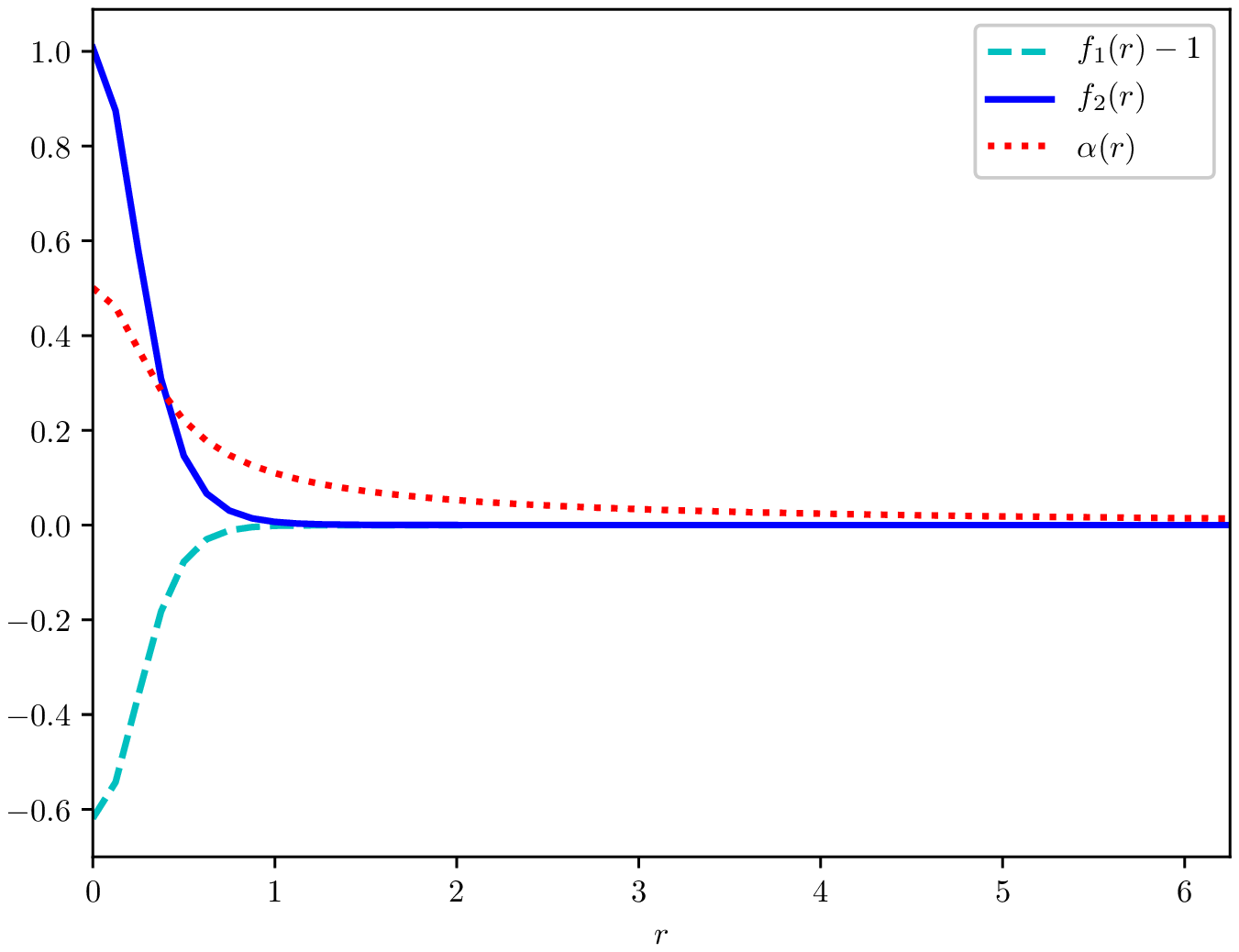}
 \caption{}
\label{fig:e10xmplef}
\end{subfigure}
\begin{subfigure}[t]{0.5\textwidth}
 \noindent\hfil\includegraphics[scale=.5]{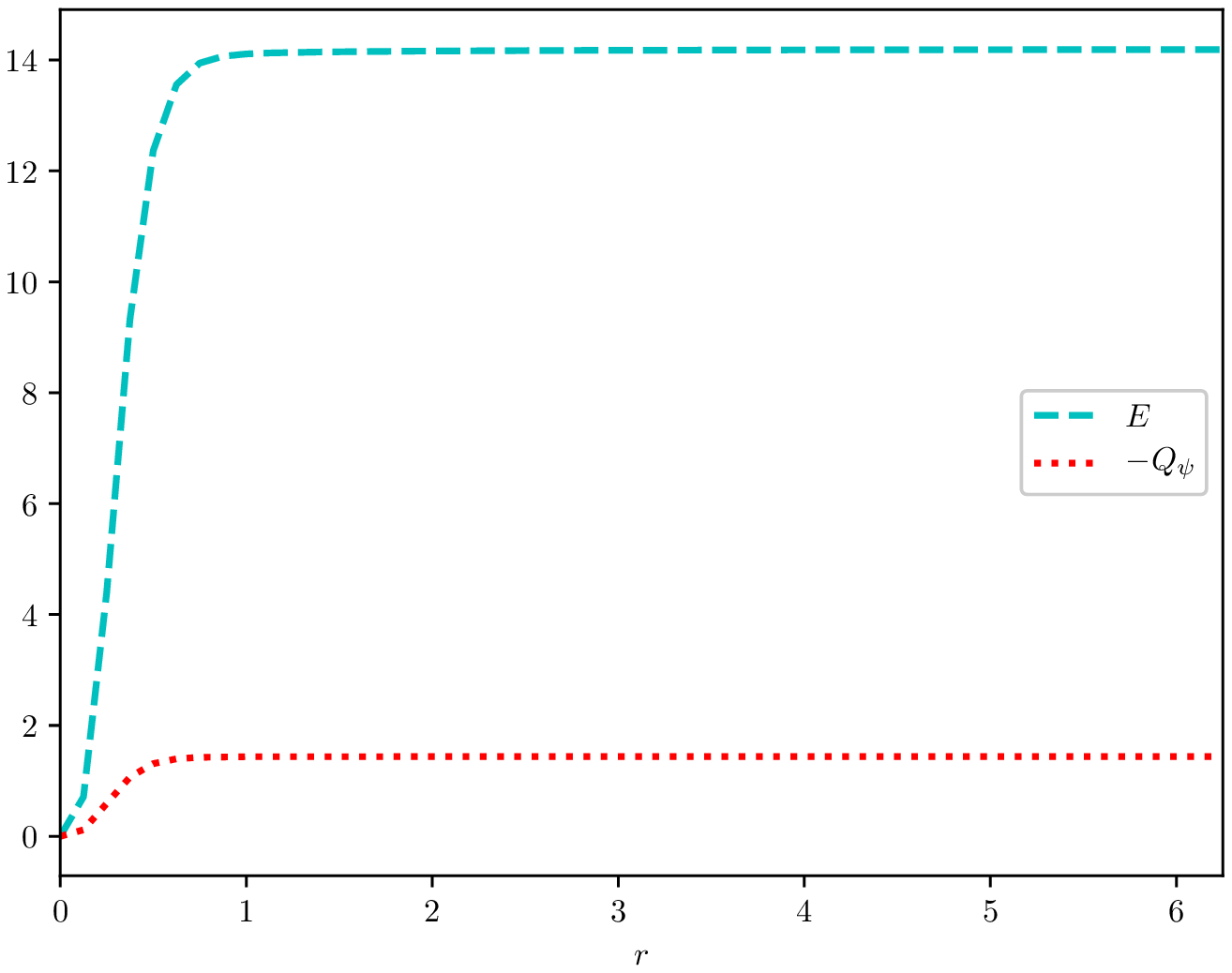}
 \caption{}
 \label{fig:e10xmplee}
\end{subfigure}
\caption{The (a) profile functions and (b) energy and charge of a Q-ball with vanishing charge $q_1=0$; $\beta_1=25$, $\beta_2=0$, $\beta'=\mu=100$, $\omega=8.8$}
\label{fig:e10xmple}
\end{figure}

The limiting cases are quite remarkable here. Considering $q=q_2/q_1>1$, the positivity condition (\ref{eq:alphapos}) is modified as $\beta_1 < \mu q_1^2 / 2$. Increasing the relative charge $q$, while keeping $\beta_1$ and $\mu$ fixed, at $q_1=\sqrt{2\beta_1/\mu}$ the positivity condition becomes violated, $\omega_{\rm min} = \omega_{\rm max}=\sqrt{\mu}$ is reached, and the solutions cease to exist.\\
For this reason the solutions of Ref.\ \cite{LeeYoon} at $q_1=0$ thus cannot be reached from the solutions considered here by increasing the charge ratio $q$ continuously; they seem to belong to a different family of solutions, with the electromagnetic field massless, and $\alpha$ approaching its limit at $r\to\infty$ as $\propto Q_\psi/r$. We have found numerical evidence (see Fig.\ \ref{fig:e10xmple}) showing that this family of solutions can also be deformed to $q_1\ne 0$, at least for small values. Their detailed investigation will be part of a further study.

\begin{figure}[h!]
\begin{subfigure}[t]{0.5\textwidth}
 \noindent\hfil\includegraphics[scale=.5]{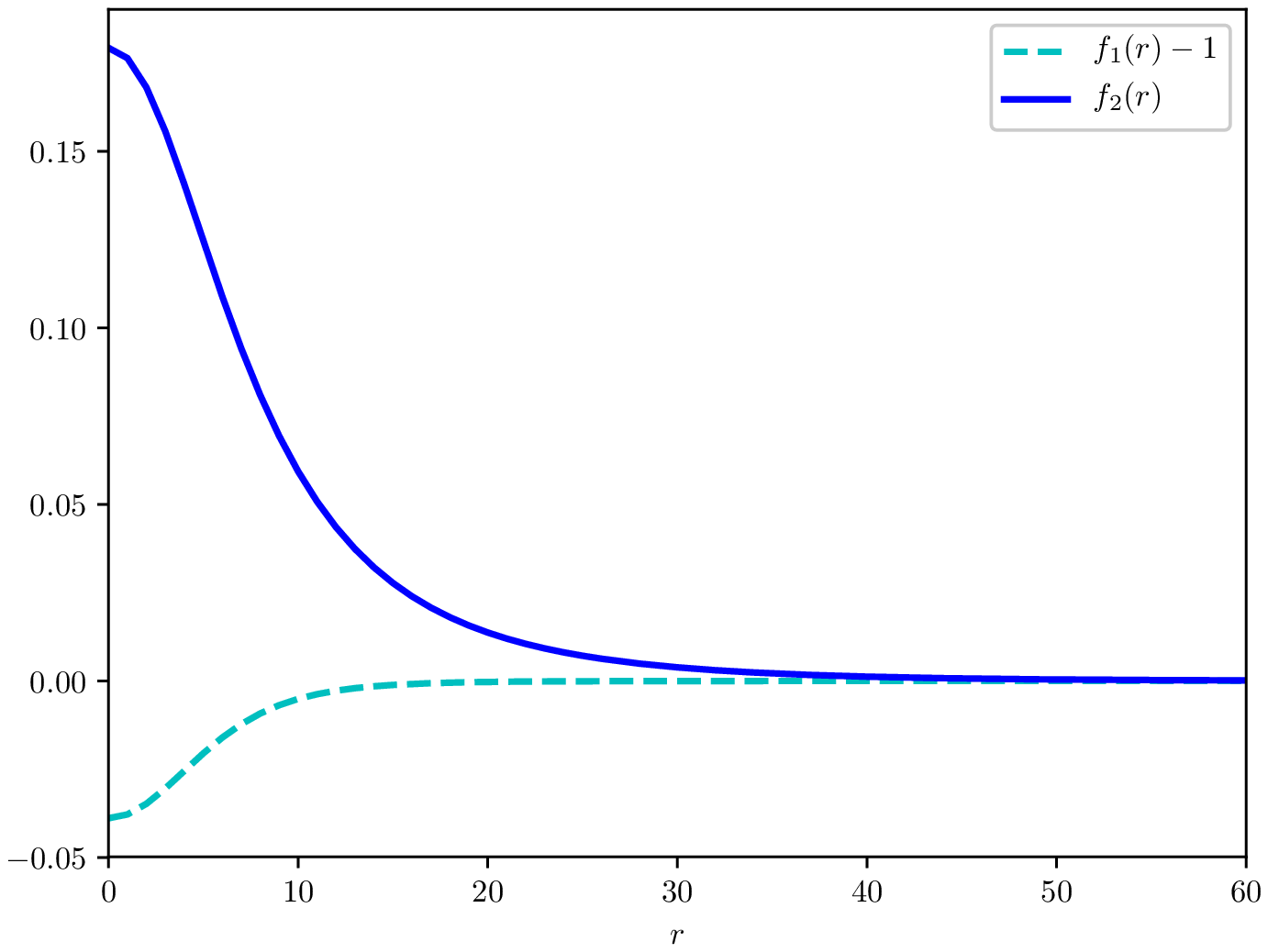}
 \caption{}
\label{fig:e20xmplef}
\end{subfigure}
\begin{subfigure}[t]{0.5\textwidth}
 \noindent\hfil\includegraphics[scale=.5]{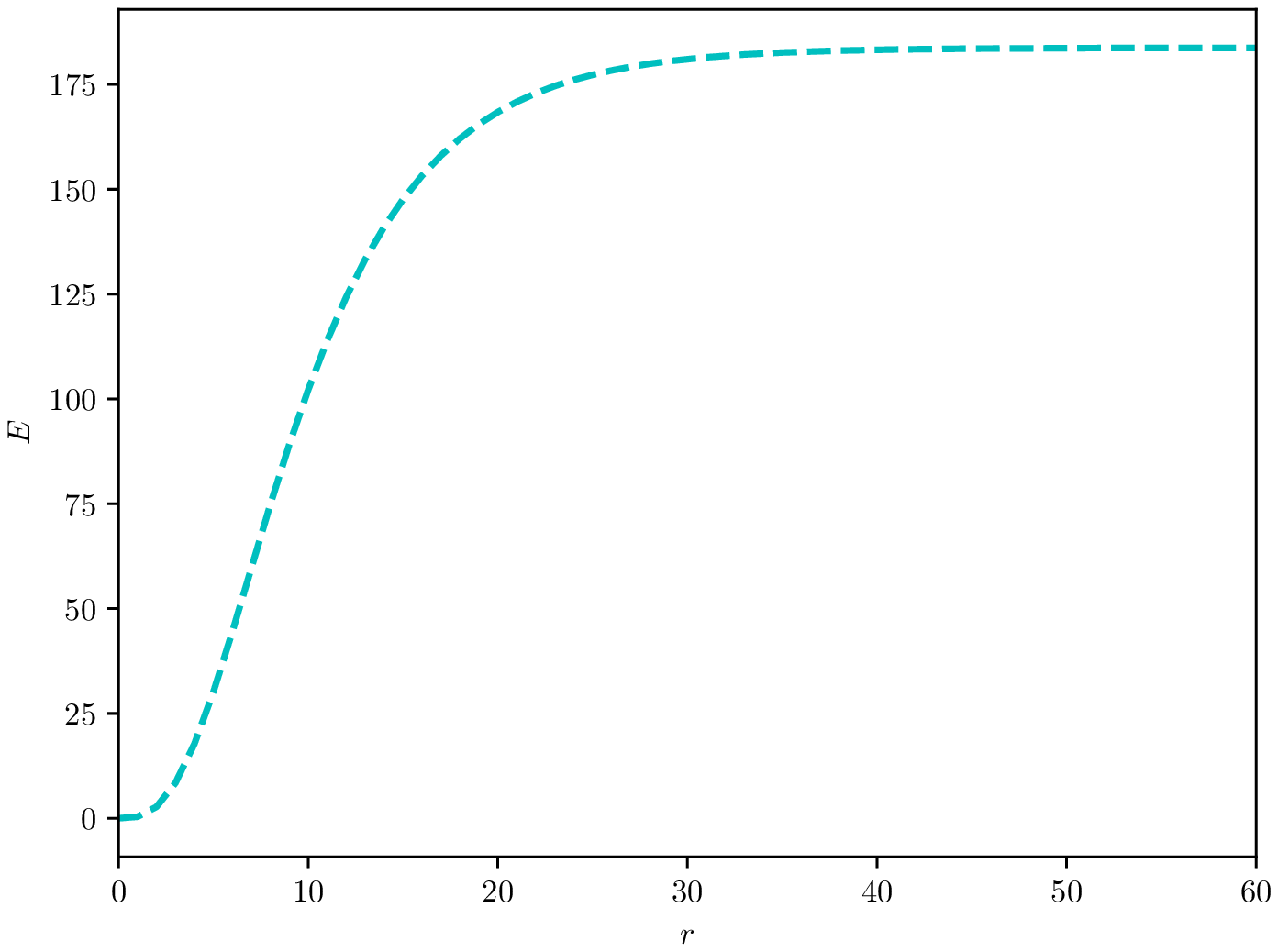}
 \caption{}
 \label{fig:e20xmplee}
\end{subfigure}
\caption{The (a) profile functions and (b) energy of a Q-ball with vanishing charge $q_2=0$; $\beta_1=0.5$, $\beta_2=0$, $\beta_{12}=\mu=1.4$, $\omega=1.180$.}
\label{fig:e20xmple}
\end{figure}

The limit $q_2\to 0$ is, on the other hand, quite simple. In this case, as $q_2$ decreases, the sum of the local charges $Q_\phi+Q_\psi$ becomes smaller, and at $q_2=0$ the purely scalar solutions of Ref.\ \cite{friedbergleesirlin} are recovered, (obviously with $\alpha=0$), see Fig.\ \ref{fig:e20xmple}.

\subsection{The limit of vanishing quartic couplings}\label{ssec:zeropot}
\begin{figure}[h!]
\begin{subfigure}[t]{0.5\textwidth}
 \noindent\hfil\includegraphics[scale=.5]{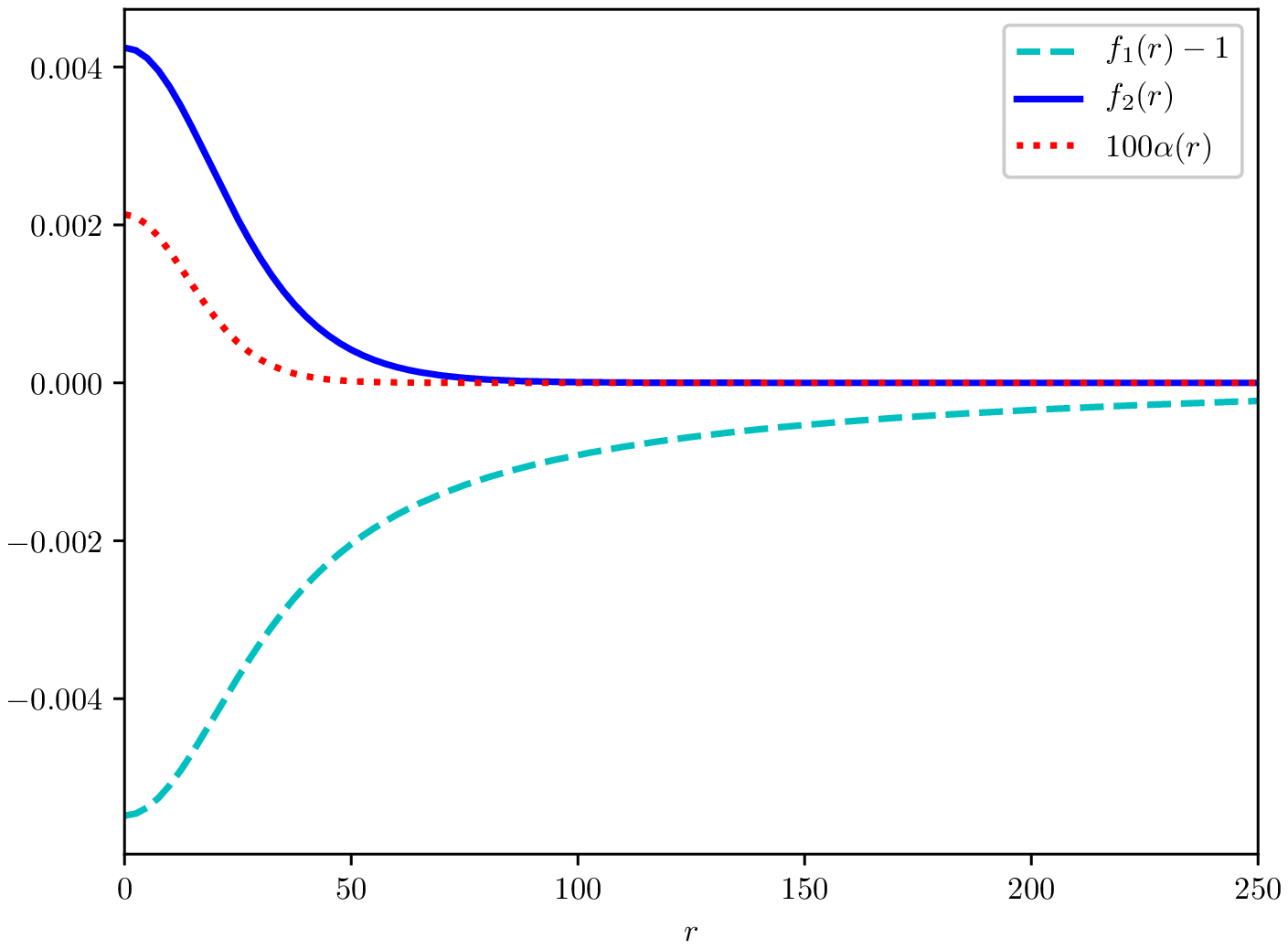}
 \caption{}
 \label{fig:b0xmplef}
\end{subfigure}
\begin{subfigure}[t]{0.5\textwidth}
 \noindent\hfil\includegraphics[scale=.5]{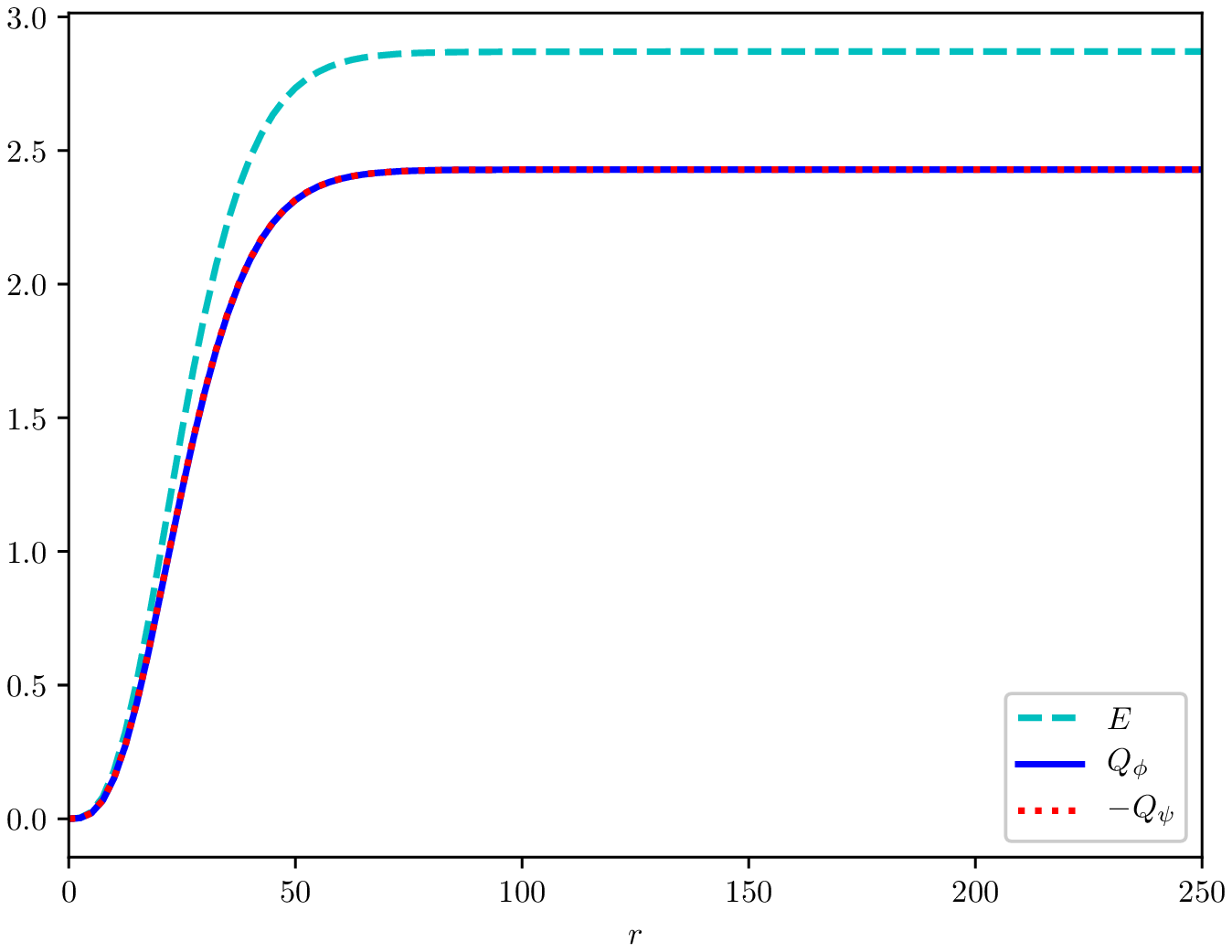}
 \caption{}
\label{fig:b0xmplee}
\end{subfigure}
\caption{(a) The profile functions of a Q-ball with vanishing quartic couplings; $\beta_1=\beta_2=0$, $\beta_{12}=\mu=1.4$, $\omega=1.180$. (b) The energy and charge distributions of the same Q-ball.}
\label{fig:b0xmple}
\end{figure}

Another interesting limiting case is $\beta_{1,2}\to 0$. For this to be a regular limit, the boundary conditions $f_1\to 1$, $f_2,\alpha \to 0$ as $r\to\infty$ are kept. We have constructed such solutions, parameterised by a frequency $0<\omega <\sqrt{\mu}$. For an example of such a solution, see Fig.\ \ref{fig:b0xmple}.

\begin{figure}[h!]
\begin{subfigure}[t]{0.5\textwidth}
 \noindent\hfil\includegraphics[scale=.5]{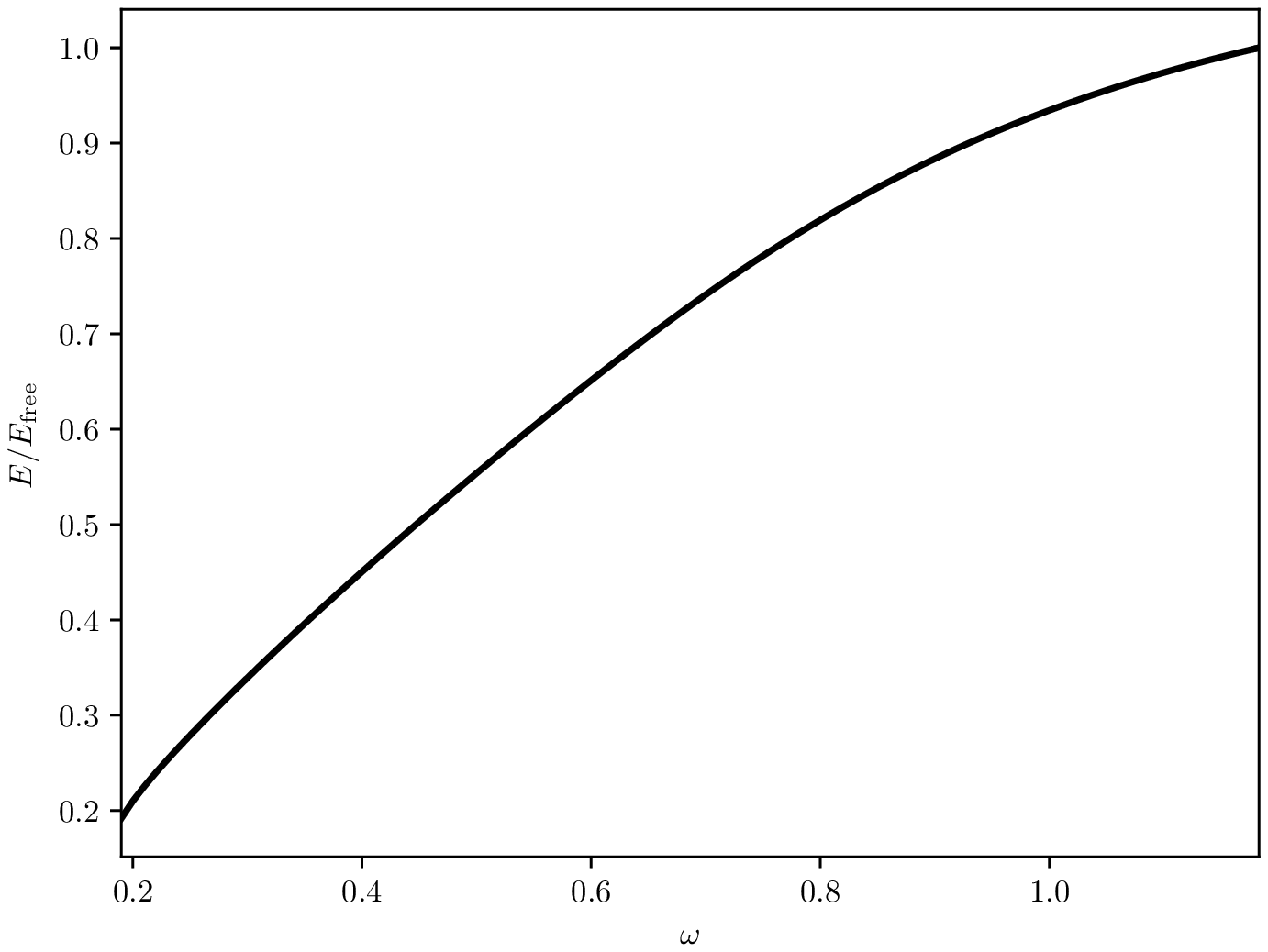}
 \caption{}
 \label{fig:stabEEomb0}
\end{subfigure}
\begin{subfigure}[t]{0.5\textwidth}
 \noindent\hfil\includegraphics[scale=.5]{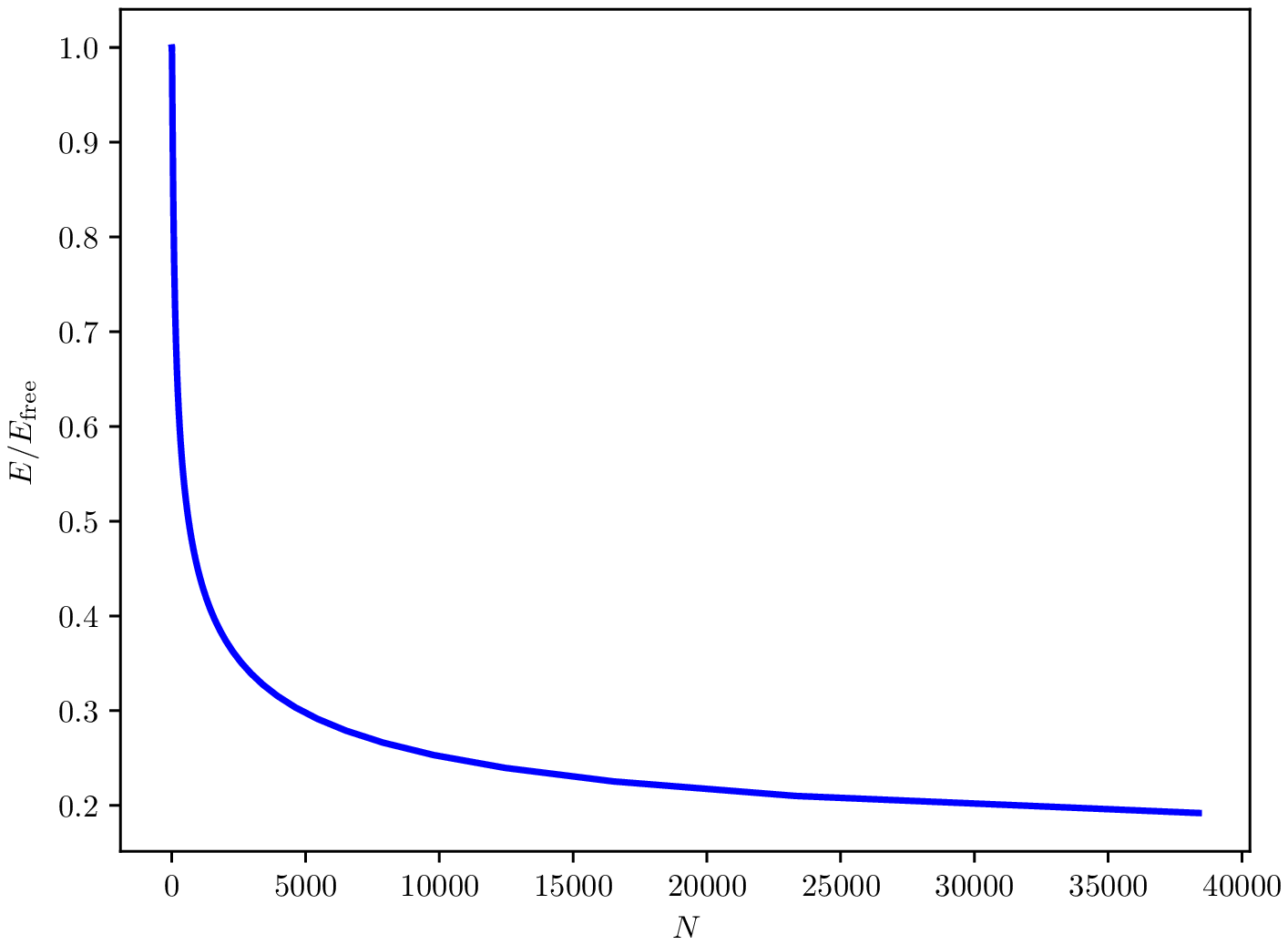}
 \caption{}
 \label{fig:stabEENb0}
\end{subfigure}
\caption{The ratio of the energy of the solutions and that of free $\psi$ particles of the same charge, for $\beta_1=0$, $\beta_2=0$, $\beta_{12}=\mu=1.4$, (a) as a function of the frequency $\omega$ and (b) of the number of $\psi$ particles in the Q-ball.}
\label{fig:stabb0}
\end{figure}

In the small frequency limit, the charges and the energy of the solutions increase, and $E/E_{\rm free}$, the total energy over the mass of free particles of type $\psi$, decreases. Also, interestingly, as $\omega$ approaches $\omega_{\rm max}$, unlike in the other cases, charge and energy do not diverge, instead, the solutions approach the vacuum. Also, instability due to $E/E_{\rm free} >1$ and the cusp on the $E/E_{\rm free}$ -- $N$ curve has not been observed here (see Fig.\ \ref{fig:stabb0}).

\section{Conclusions}
In the present paper, we have extended the investigations of Refs.\ \cite{ishiharaogawasoln, ishiharaogawasoln2} to the case of the most general U(1)$\times$U(1) potential, when both scalar fields of the model are self-interacting.
We have shown, that the perfect global charge screening is a consequence of Gauss' theorem. The local charge screening,  i.e., a high degree of cancellation between the charge densities of the scalar fields in the outer region of the Q-ball, as found in Refs.\ \cite{ishiharaogawascreen, ishiharaogawasoln, ishiharaogawasoln2} is still present. 
The main effect of the self-interaction of the second scalar  field
 is an increase of the energy, of the charge, and in most cases, a
smaller domain of existence.
We have also exhibited a new family of solutions where both quartic scalar self-interactions terms are absent, and found that some of their properties differ qualitatively from the generic case.

\paragraph{Acknowledgements} We thank Prof.~Hideki Ishihara for correspondence concerning the numerical methods used in Refs.\ \cite{ishiharaogawasoln, ishiharaogawasoln2}.
We acknowledge the support of  the Spanish Ministerio de Ciencia, Innovación y Universidades (Grant No.\ PCI2018-092896) and the EU (QuantERA CEBBEC).

\def\refttl#1{{\sl ``#1''}, }%

\appendix
\newpage\clearpage
\section{Supplementary material: numerical data}\label{app:num}
In Tables\ \ref{tab:om1175}-\ref{tab:om1181} we have collected the data of a number of Q-ball solutions. The parameters $f_{1,2}(0)$ and $\alpha(0)$ are given for reproducibility. The energy and the charges are defined as the integrals of the respective densities from 0 to 200. 

\begin{figure}[h!]
\begin{subfigure}[t]{0.5\textwidth}
 \noindent\hfil\includegraphics[scale=.5]{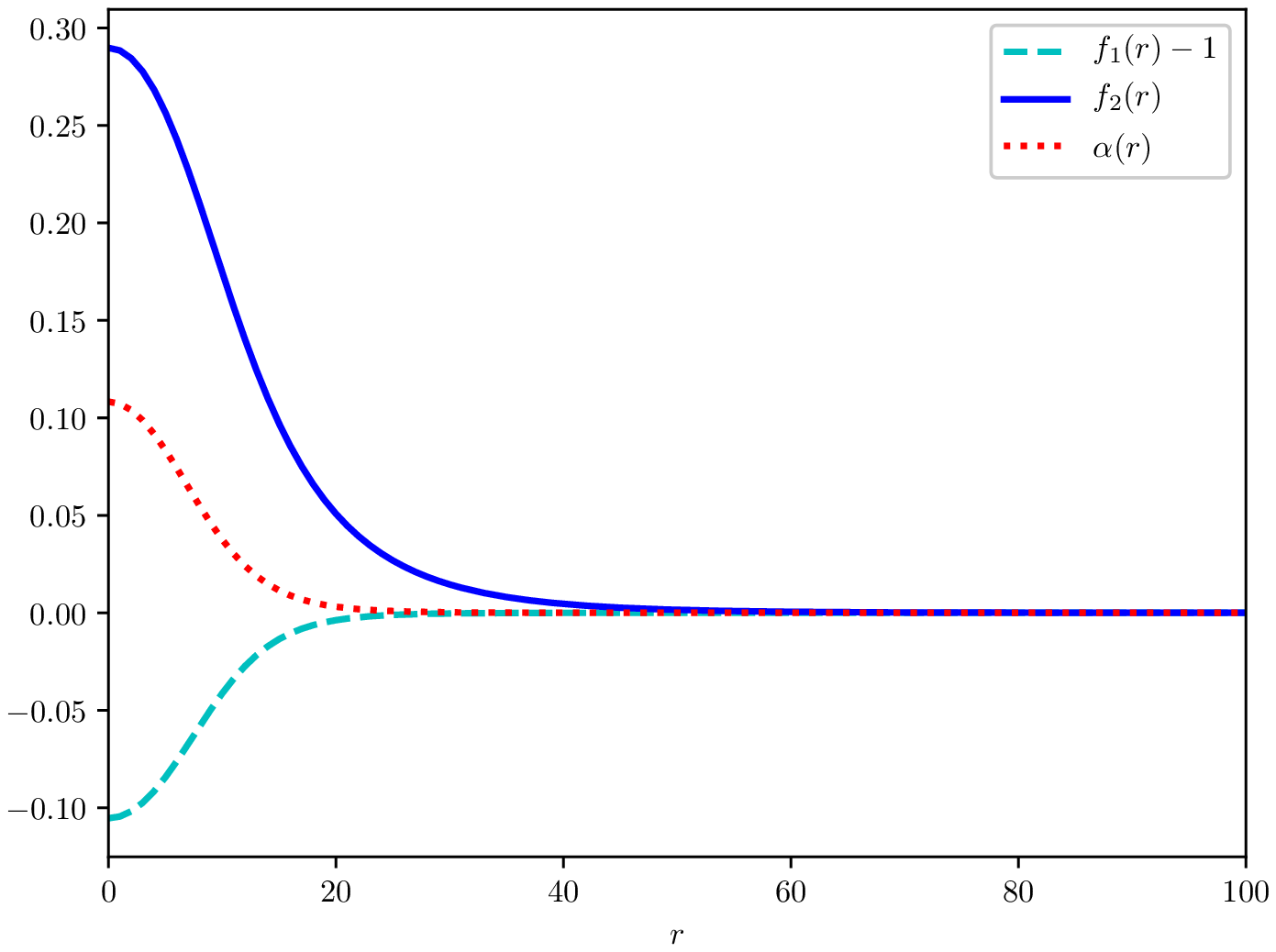}
 \caption{}
 \label{fig:xamplef}
\end{subfigure}
\begin{subfigure}[t]{0.5\textwidth}
 \noindent\hfil\includegraphics[scale=.5]{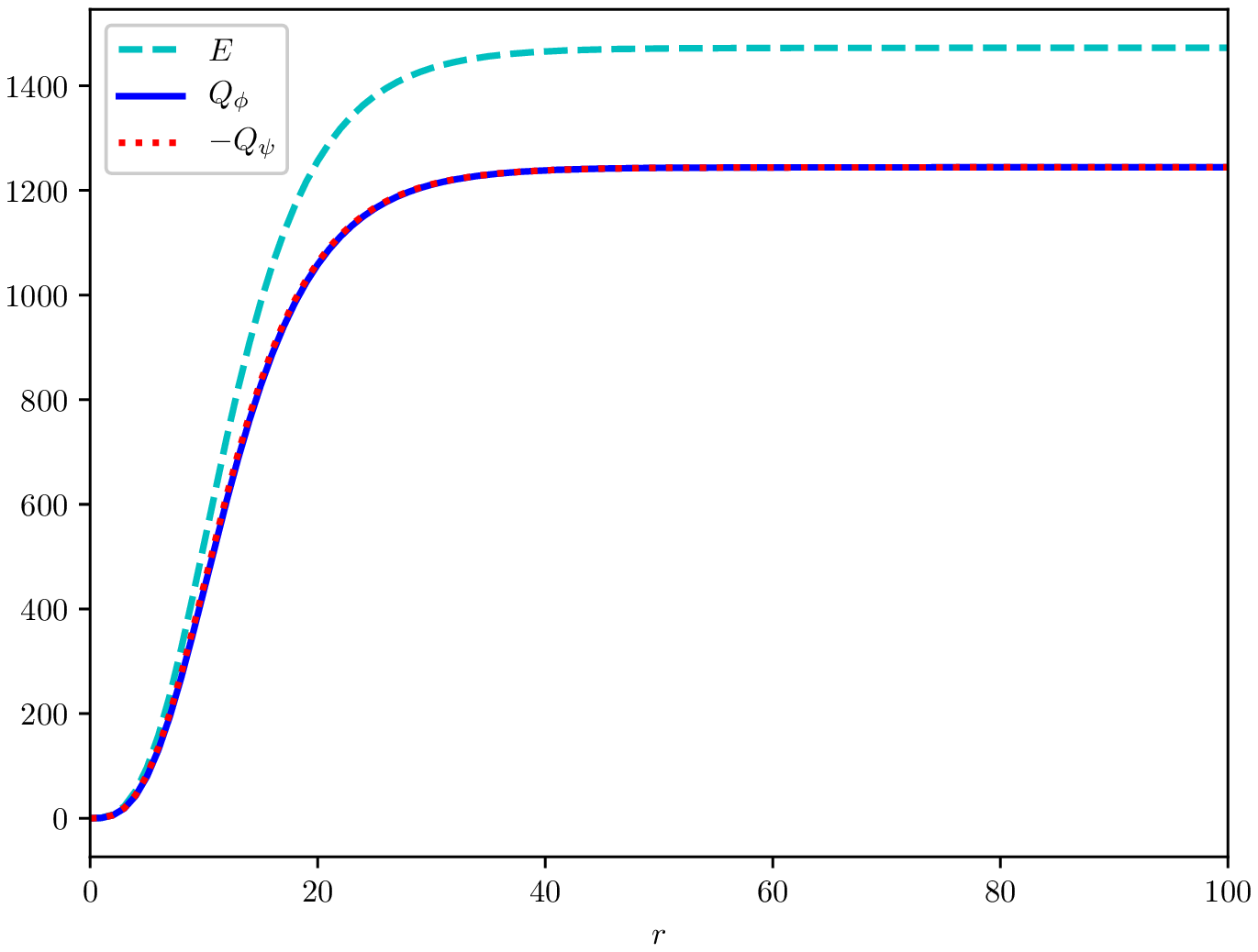}
 \caption{}
 \label{fig:xamplee}
\end{subfigure}
\caption{(a) The profile functions of a typical Q-ball; $\beta_1=0.5$, $\beta_{12}=\mu=1.4$, $\beta_2=0$, $\omega=0.180$. (b) The energy and charge distributions of the same Q-ball.}
\label{fig:xample}
\end{figure}

\begin{figure}[h!]
\begin{subfigure}[t]{0.5\textwidth}
 \noindent\hfil\includegraphics[scale=.5]{omd1.eps}
 \caption{$\omega=1.18$}
\end{subfigure}
\begin{subfigure}[t]{0.5\textwidth}
 \noindent\hfil\includegraphics[scale=.5]{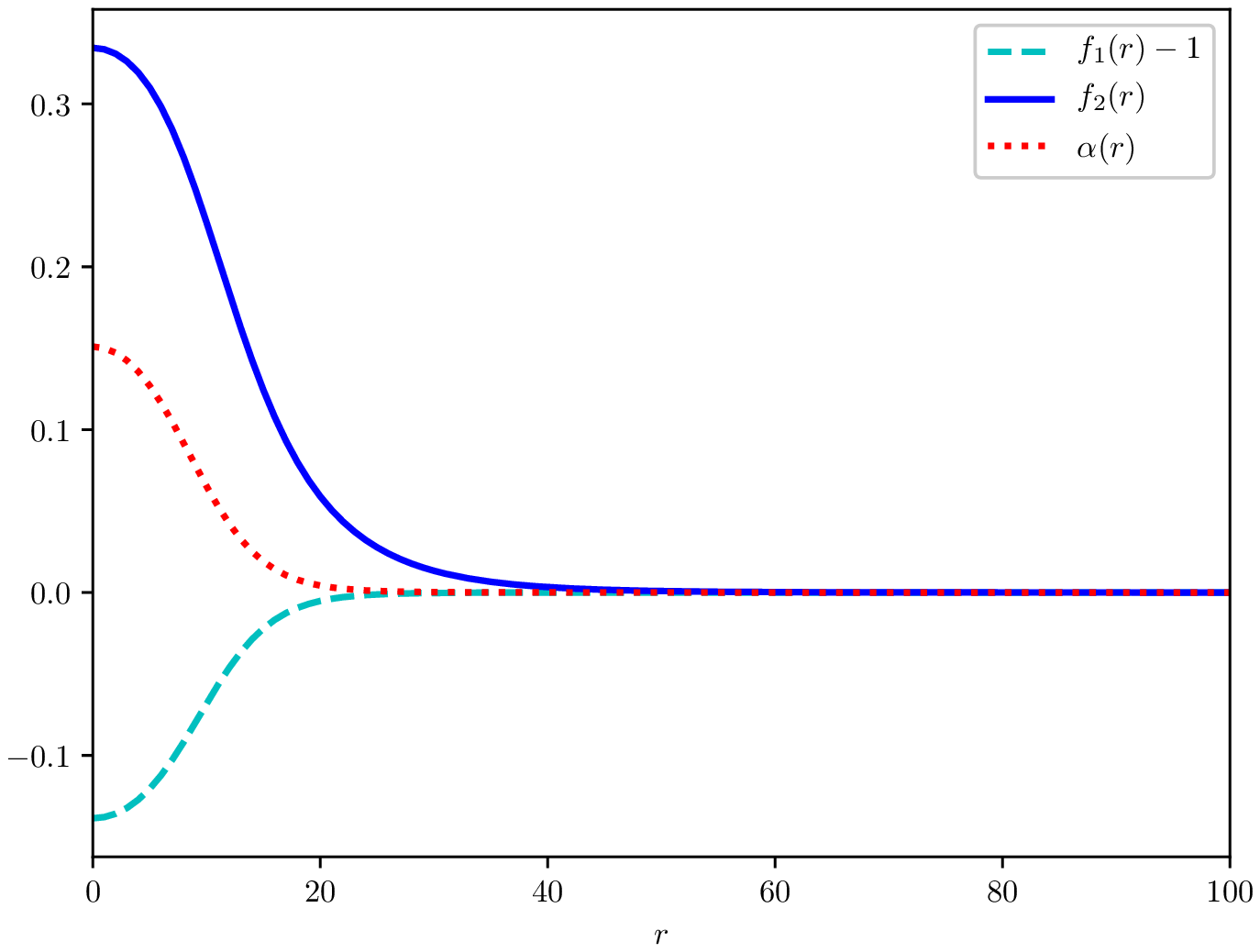}
 \caption{$\omega=1.178$}
\end{subfigure}
 \begin{subfigure}[t]{0.5\textwidth}
 \noindent\hfil\includegraphics[scale=.5]{omd3.eps}
 \caption{$\omega=1.176$}
\end{subfigure}
\begin{subfigure}[t]{0.5\textwidth}
 \noindent\hfil\includegraphics[scale=.5]{omd4.eps}
 \caption{$\omega=1.174$}
\end{subfigure}
\caption{The frequency dependence of the solutions, $\beta_1=0.5$, $\beta_{12}=\mu=1.4$, $\beta_2=0$.}
\label{fig:4omd}
\end{figure}

\begin{figure}[h!]
\begin{subfigure}[t]{0.5\textwidth}
 \noindent\hfil\includegraphics[scale=.5]{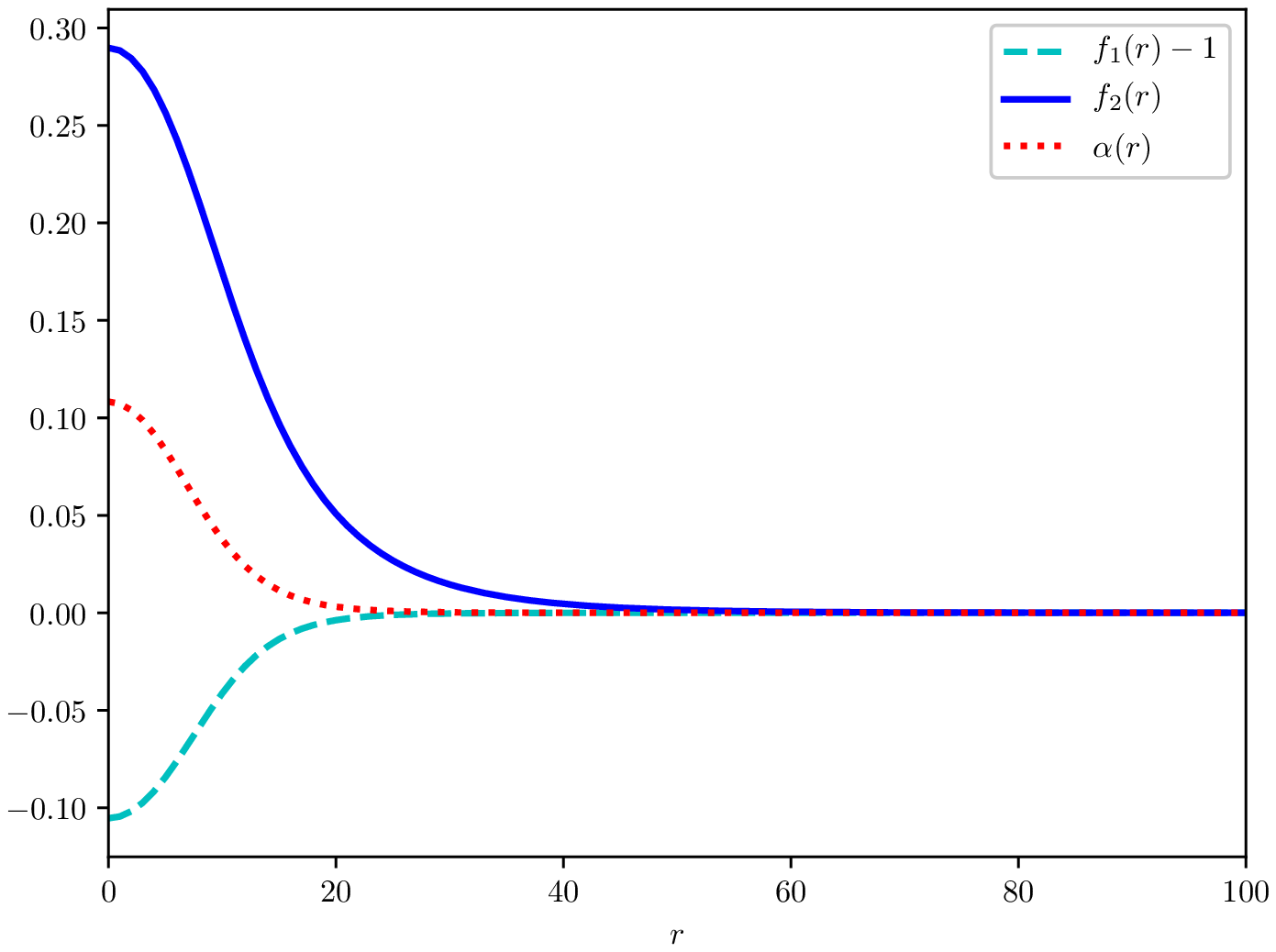}
 \caption{$\mu=1.4$}
\end{subfigure}
\begin{subfigure}[t]{0.5\textwidth}
 \noindent\hfil\includegraphics[scale=.5]{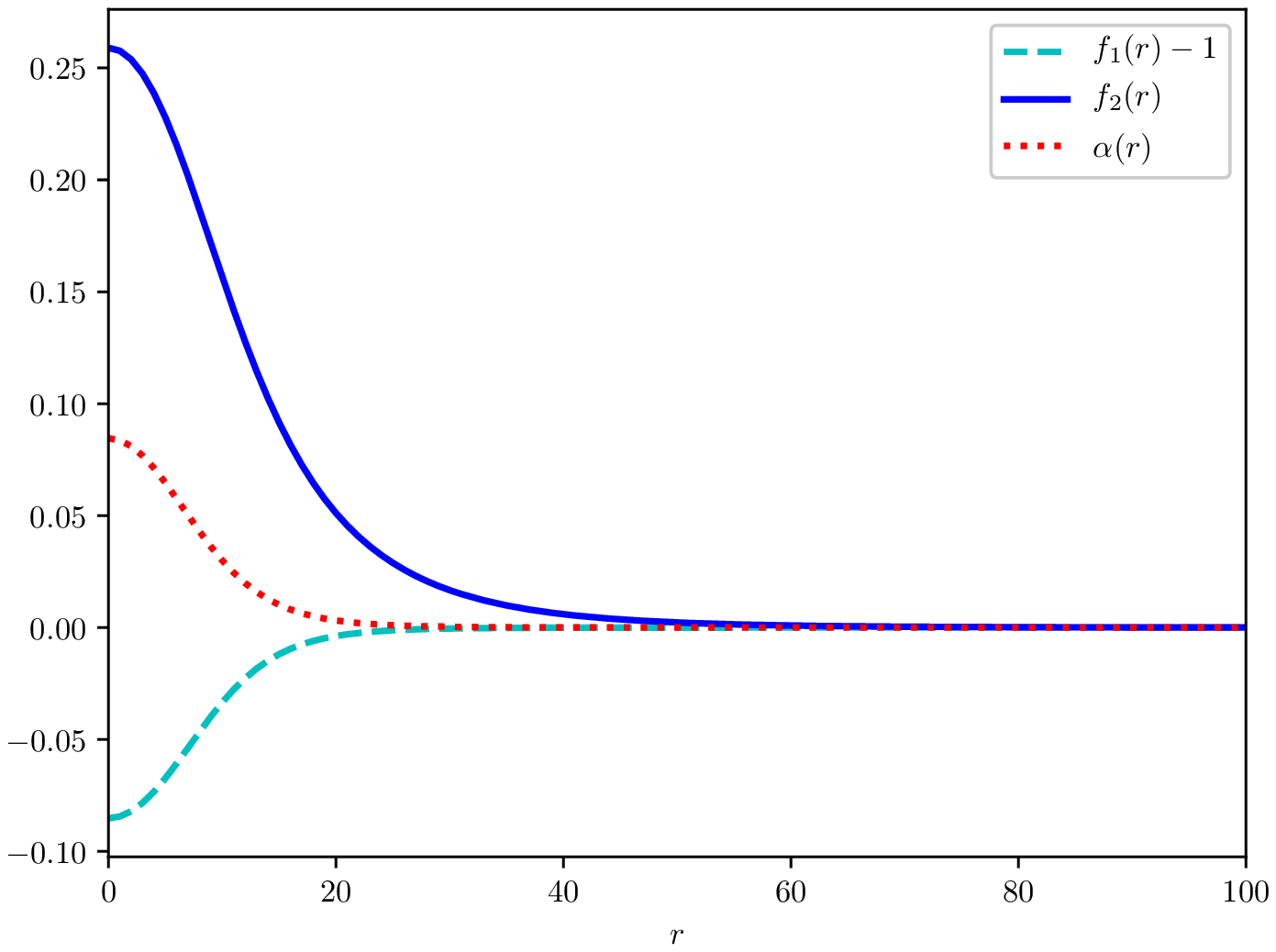}
 \caption{$\mu=1.398$}
\end{subfigure}
\begin{subfigure}[t]{0.5\textwidth}
 \noindent\hfil\includegraphics[scale=.5]{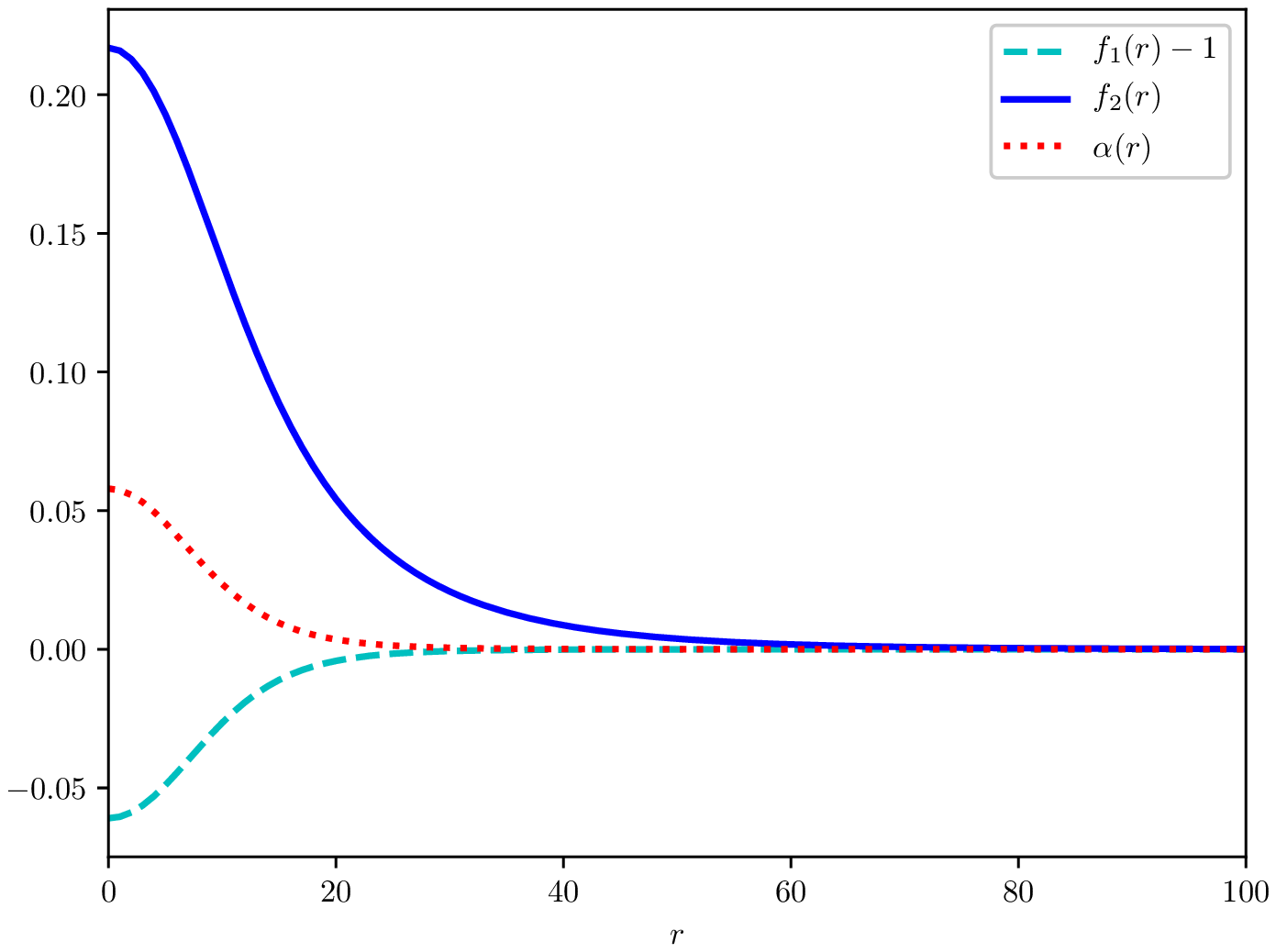}
 \caption{$\mu=1.396$}
\end{subfigure}
\begin{subfigure}[t]{0.5\textwidth}
 \noindent\hfil\includegraphics[scale=.5]{mud4.eps}
 \caption{$\mu=1.394$}
\end{subfigure}
%
\caption{The dependence of the solutions on the parameter $\mu$ approaching the lower limit $\mu=\omega^2$, $\beta_1=0.5$, $\beta_2=0$, $\beta_{12}=1.4$, $\omega=1.18$.}
\label{fig:4mud}
\end{figure}

\begin{figure}[h!]
\begin{subfigure}[t]{0.5\textwidth}
 \noindent\hfil\includegraphics[scale=.5]{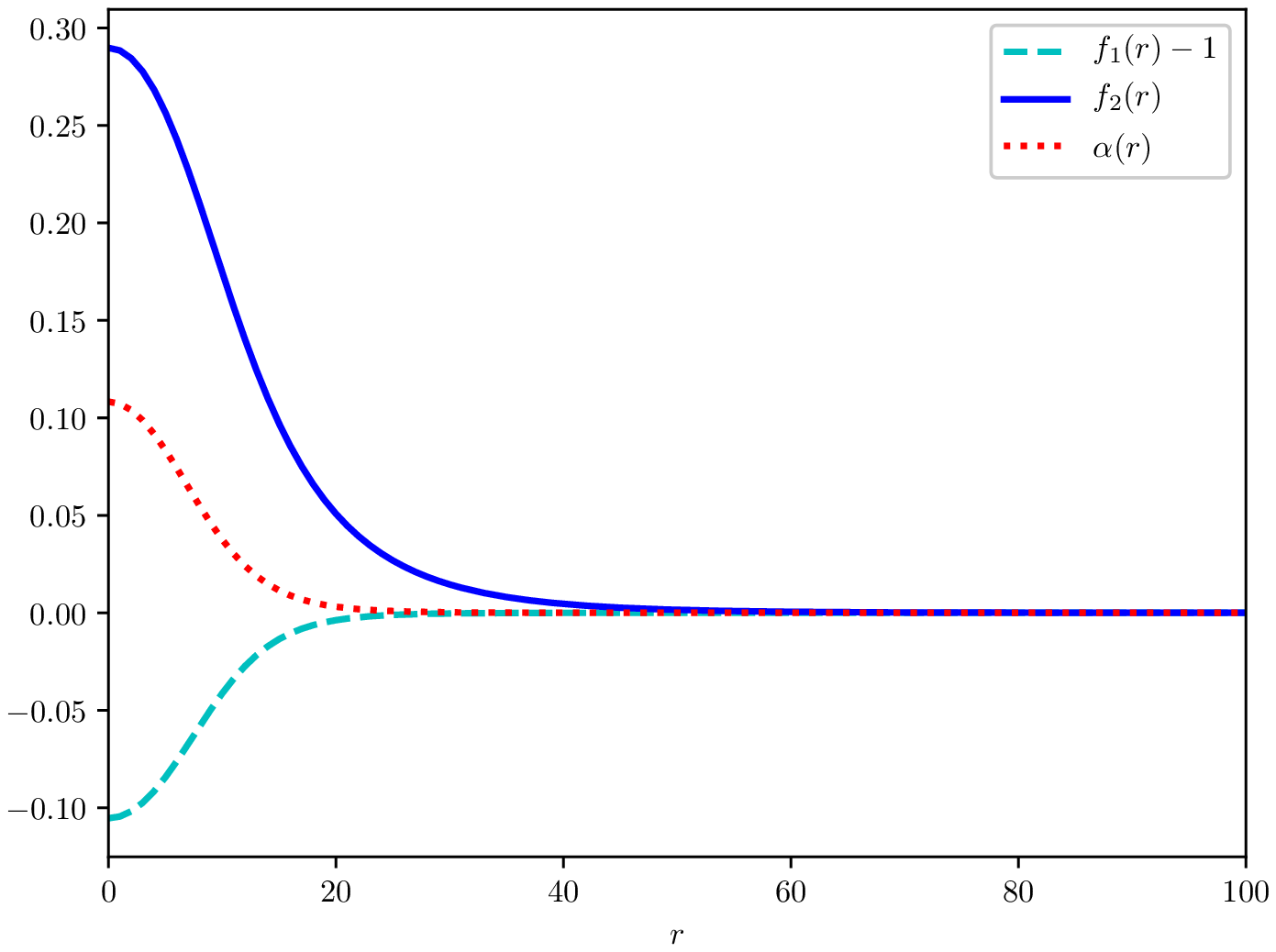}
 \caption{$\mu=1.4$}
\end{subfigure}
\begin{subfigure}[t]{0.5\textwidth}
 \noindent\hfil\includegraphics[scale=.5]{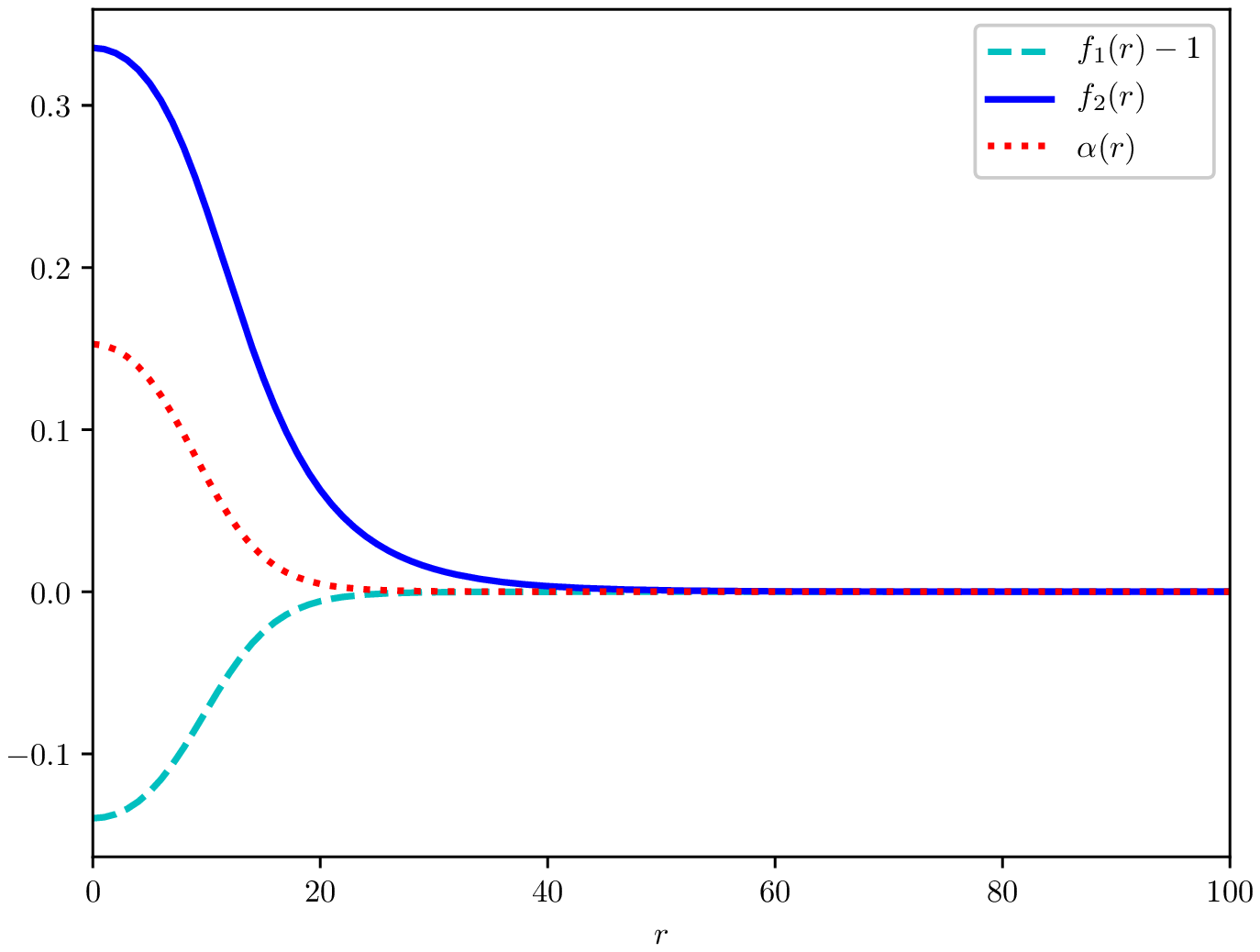}
 \caption{$\mu=1.405$}
\end{subfigure}
 \begin{subfigure}[t]{0.5\textwidth}
 \noindent\hfil\includegraphics[scale=.5]{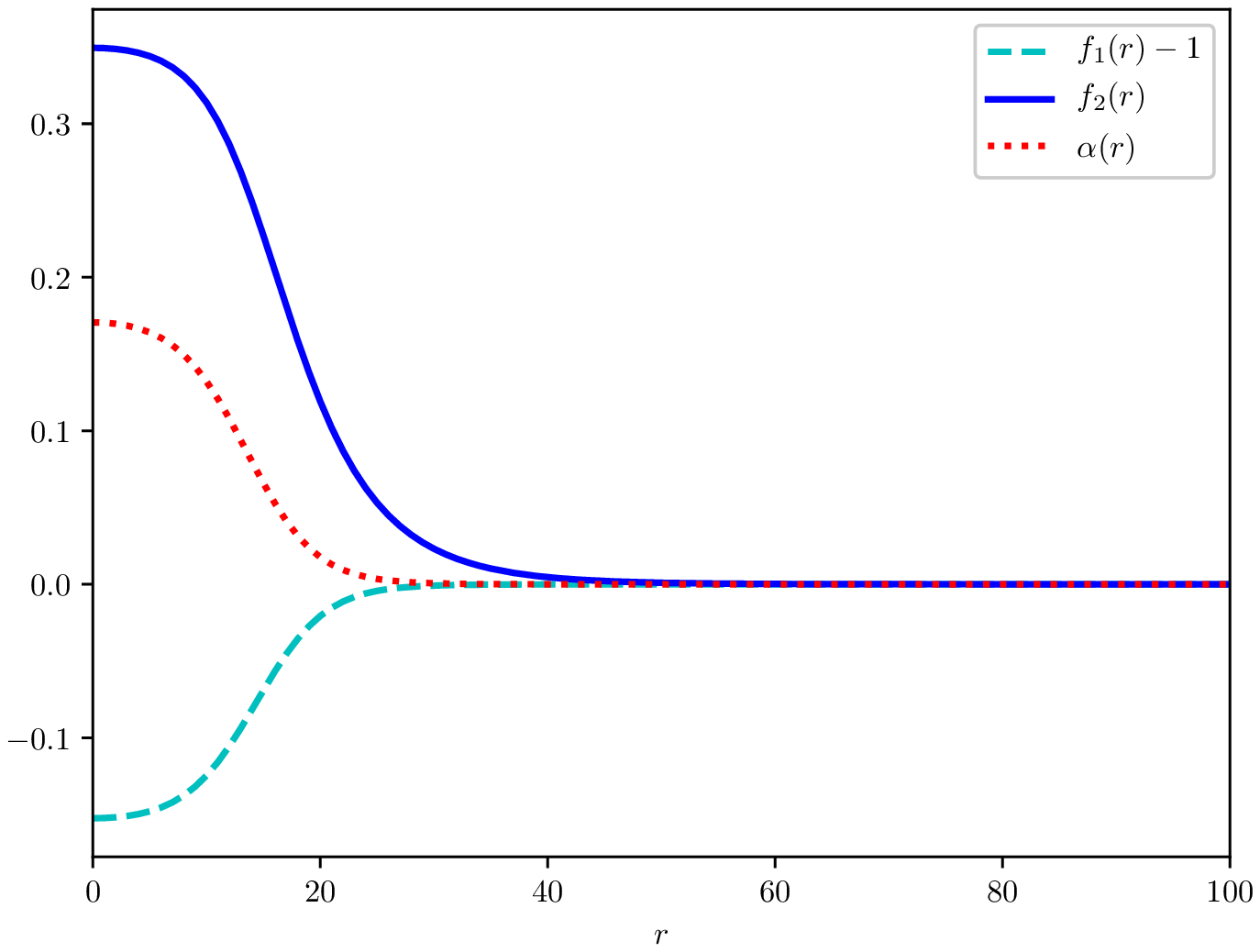}
 \caption{$\mu=1.41$}
\end{subfigure}
\begin{subfigure}[t]{0.5\textwidth}
 \noindent\hfil\includegraphics[scale=.5]{muu4.eps}
 \caption{$\mu=1.413$}
\end{subfigure}
\caption{The dependence of the solutions on the parameter $\mu$ approaching the upper limit $\mu=\mu_{\rm max}$, $\beta_1=0.5$, $\beta_2=0$, $\beta_{12}=1.4$, $\omega=1.18$.}
\label{fig:4muu}
\end{figure}

In addition, we present some more figures depiciting the chage of profile functions with various parameters approaching their extreme values. Fig.\ \ref{fig:xample} is the reconstruction of a solution considered in Refs.\ \cite{ishiharaogawasoln, ishiharaogawasoln2}. Fig.\ \ref{fig:4omd} shows how the  Q-ball deforms as $\omega\to \omega_{\rm min}$, whereas Fig.\ \ref{fig:4omd} what happens as $\omega\to\omega_{\rm max}$. Similarly, Fig.\ \ref{fig:4mud} shows the limit $\mu\to \mu_{\rm min}=\omega^2$, and Fig.\ \ref{fig:4muu} the one $\mu\to\mu_{\rm max}$.

\begin{table}[h!]
 \begin{center}
  \begin{tabular}{|c|c c c|c|c c|}
   \hline
   $\beta_2$ & $1-f_1(0)$ & $f_2(0)$ & $\alpha(0)$ & $E$ & $Q_{\phi}$ & $Q_{\psi}$ \\
   \hline\hline
0.000  & 0.1568 & 0.3549 & 0.1762   &        6018.67  &       5108.15  &      -5108.15 \\
0.025  & 0.1536 & 0.3501 & 0.1713   &        7595.64  &       6448.36  &      -6448.35 \\
0.050  & 0.1501 & 0.3450 & 0.1660   &        9903.31  &       8409.84  &      -8409.84 \\
0.100  & 0.1421 & 0.3337 & 0.1544   &       19325.37  &      16420.36  &      -16420.34 \\
0.125  & 0.1377 & 0.3276 & 0.1482   &       29910.14  &      25421.23  &      -25421.22 \\
   \hline
  \end{tabular}
 \end{center}
 \caption{Properties of Q-balls for $\beta_1=0.5$, $\beta_{12}=\mu=1.4$ and $\omega=1.175$}
 \label{tab:om1175}
\end{table}

\begin{table}[h!]
 \begin{center}
  \begin{tabular}{|c|c c c|c|c c|}
   \hline
   $\beta_2$ & $1-f_1(0)$ & $f_2(0)$ & $\alpha(0)$ & $E$ & $Q_{\phi}$ & $Q_{\psi}$ \\
   \hline\hline
0.00 &   0.1484 & 0.3461 & 0.1645  &         2895.20  &       2451.72  &      -2451.72 \\
0.05 &   0.1458 & 0.3411 & 0.1605  &         3897.96  &       3302.28  &      -3302.28 \\
0.10 &   0.1419 & 0.3346 & 0.1546  &         5611.35  &       4755.95  &      -4755.95 \\
0.15 &   0.1365 & 0.3264 & 0.1469  &         8906.73  &       7552.49  &      -7552.49 \\
0.20 &   0.1295 & 0.3163 & 0.1372  &        16497.58  &      13995.90  &     -13995.89 \\
0.25 &   0.1210 & 0.3042 & 0.1259  &        40308.79  &      34212.94  &      -34212.92 \\
   \hline
  \end{tabular}
 \end{center}
 \caption{Properties of Q-balls for $\beta_1=0.5$, $\beta_{12}=\mu=1.4$ and $\omega=1.177$}
 \label{tab:om1177}
\end{table}

\newpage\clearpage

\begin{table}[h!]
 \begin{center}
  \begin{tabular}{|c|c c c|c|c c|}
   \hline
   $\beta_2$ & $1-f_1(0)$ & $f_2(0)$ & $\alpha(0)$ & $E$ & $Q_{\phi}$ & $Q_{\psi}$ \\
   \hline\hline
0.00 &   0.1243 & 0.3161 & 0.1321 &     1751.16 &      1480.43 &   -1480.43 \\
0.10 &   0.1257 & 0.3154 & 0.1333 &     2595.06 &      2195.00 &   -2195.00 \\
0.20 &   0.1237 & 0.3100 & 0.1301 &     4516.48 &      3822.56 &   -3822.56 \\
0.20 &   0.1237 & 0.3100 & 0.1301 &     4516.48 &      3822.56 &   -3822.56 \\
0.30 &   0.1161 & 0.2977 & 0.1199 &    10778.90 &      9129.28 &   -9129.28 \\
0.40 &   0.1011 & 0.2755 & 0.1012 &    60271.29 &     51086.82 &  -51086.78 \\
   \hline
  \end{tabular}
 \end{center}
 \caption{Properties of Q-balls for $\beta_1=0.5$, $\beta_{12}=\mu=1.4$ and $\omega=1.179$}
 \label{tab:om1179}
\end{table}

\begin{table}[h!]
 \begin{center}
  \begin{tabular}{|c|c c c|c|c c|}
   \hline
   $\beta_2$ & $1-f_1(0)$ & $f_2(0)$ & $\alpha(0)$ & $E$ & $Q_{\phi}$ & $Q_{\psi}$ \\
   \hline\hline
0.00 & 0.0813 & 0.2527 & 0.0802 &    1320.49 &    1115.40 &   -1115.40 \\
0.10 & 0.0857 & 0.2585 & 0.0850 &    1655.48 &    1398.62 &   -1398.62 \\
0.20 & 0.0899 & 0.2635 & 0.0894 &    2207.55 &    1865.46 &   -1865.46 \\
0.20 & 0.0899 & 0.2635 & 0.0894 &    2207.55 &    1865.46 &   -1865.46 \\
0.30 & 0.0930 & 0.2667 & 0.0926 &    3250.65 &    2747.74 &   -2747.74 \\
0.40 & 0.0936 & 0.2659 & 0.0929 &    5717.61 &    4834.86 &   -4834.86 \\
0.50 & 0.0886 & 0.2570 & 0.0869 &   14751.35 &   12479.75 &  -12479.75 \\
   \hline
  \end{tabular}
 \end{center}
 \caption{Properties of Q-balls for $\beta_1=0.5$, $\beta_{12}=\mu=1.4$ and $\omega=1.181$}
 \label{tab:om1181}
\end{table}

A typical solution for $\beta_2$ is shown in Fig.\ \ref{fig:xample}, and the energy and charge distributions of the same solution are shown in Fig.\ \ref{fig:xamplee}. This solution is one of the family constructed in Refs.\ \cite{ishiharaogawasoln, ishiharaogawasoln2}.

\begin{figure}[h!]
\begin{subfigure}[t]{0.5\textwidth}
 \noindent\hfil\includegraphics[scale=.5]{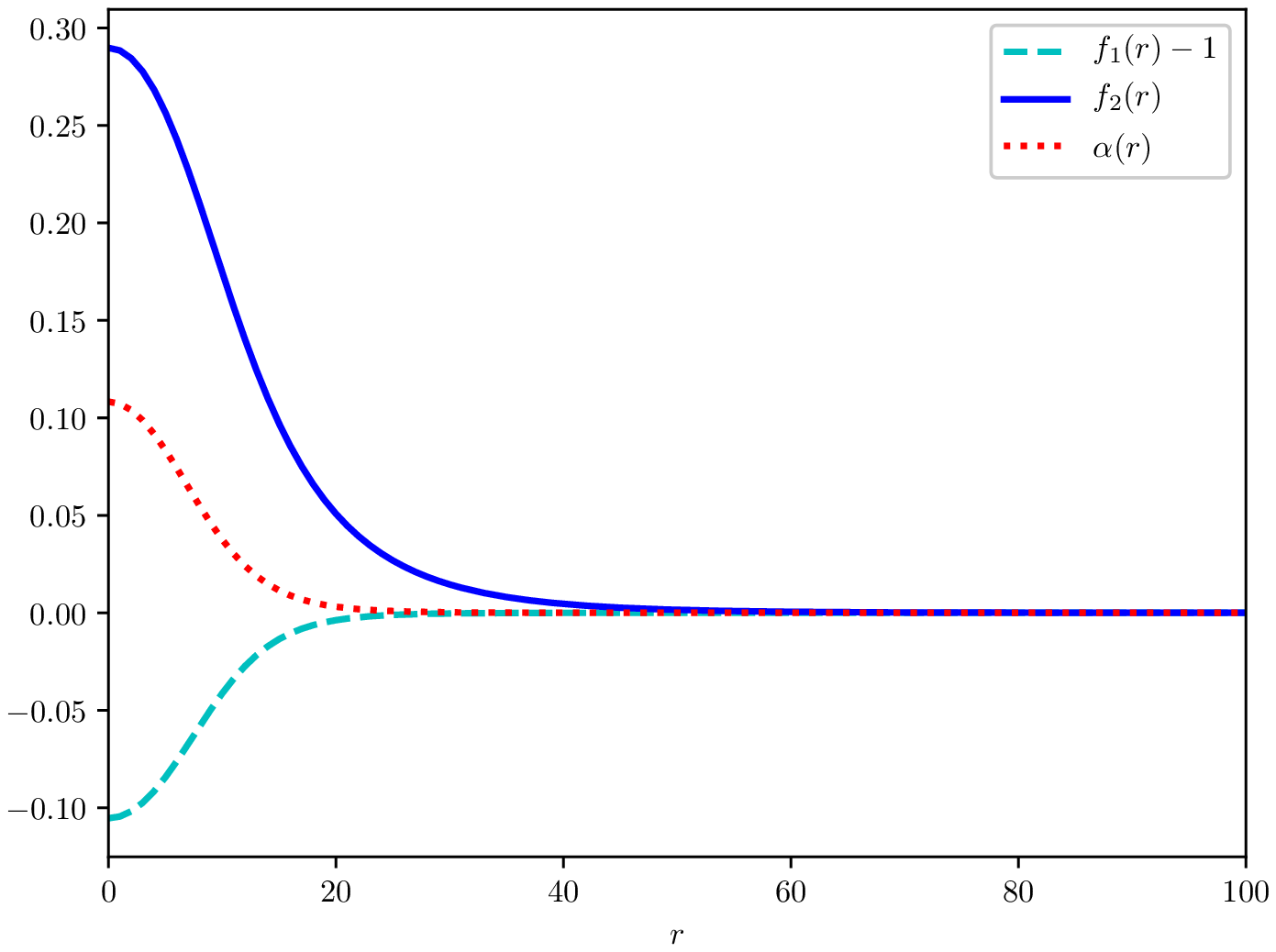}
 \caption{$\omega=1.18$}
\end{subfigure}
\begin{subfigure}[t]{0.5\textwidth}
 \noindent\hfil\includegraphics[scale=.5]{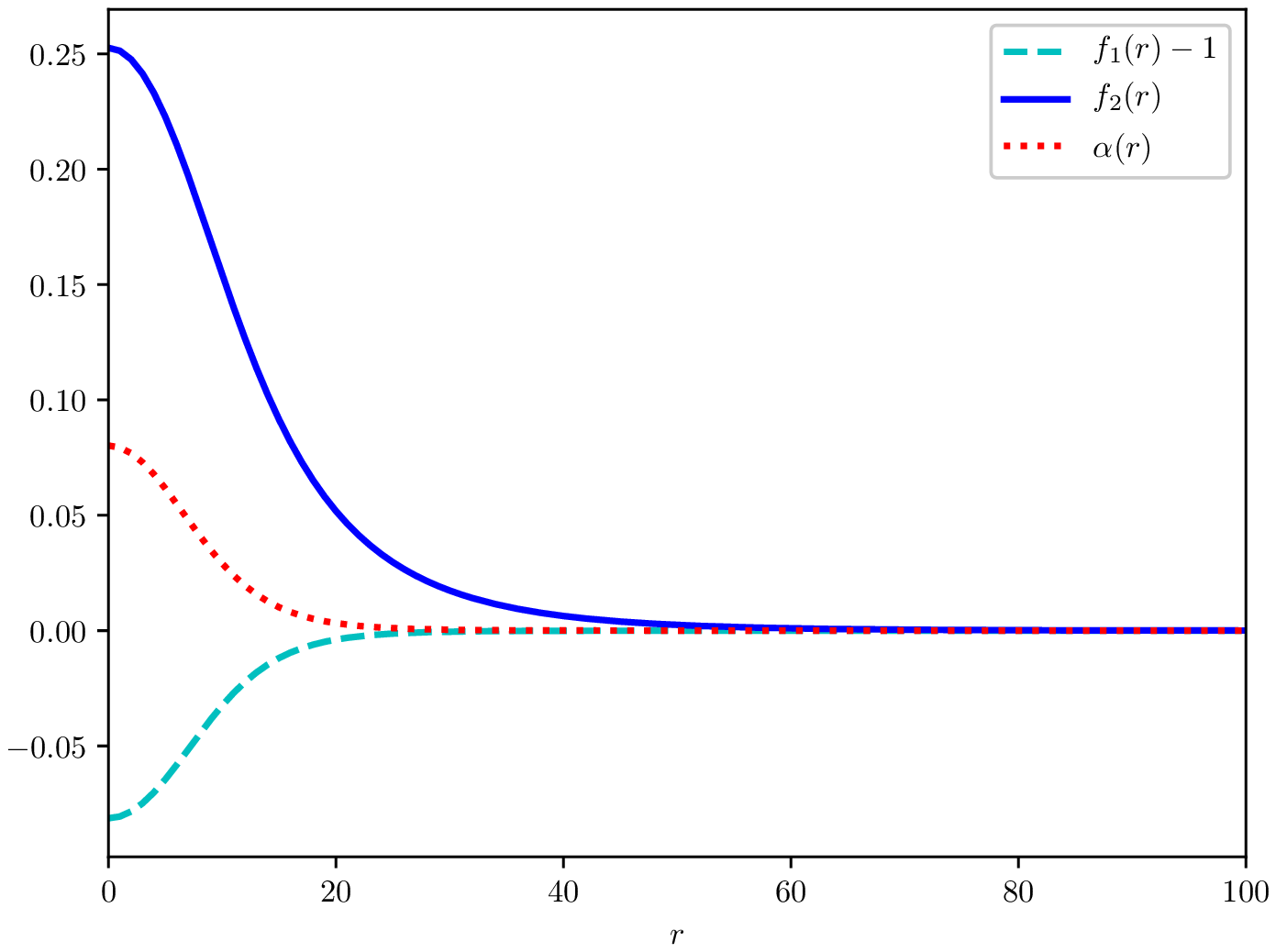}
 \caption{$\omega=1.181$}
\end{subfigure}
\begin{subfigure}[t]{0.5\textwidth}
 \noindent\hfil\includegraphics[scale=.5]{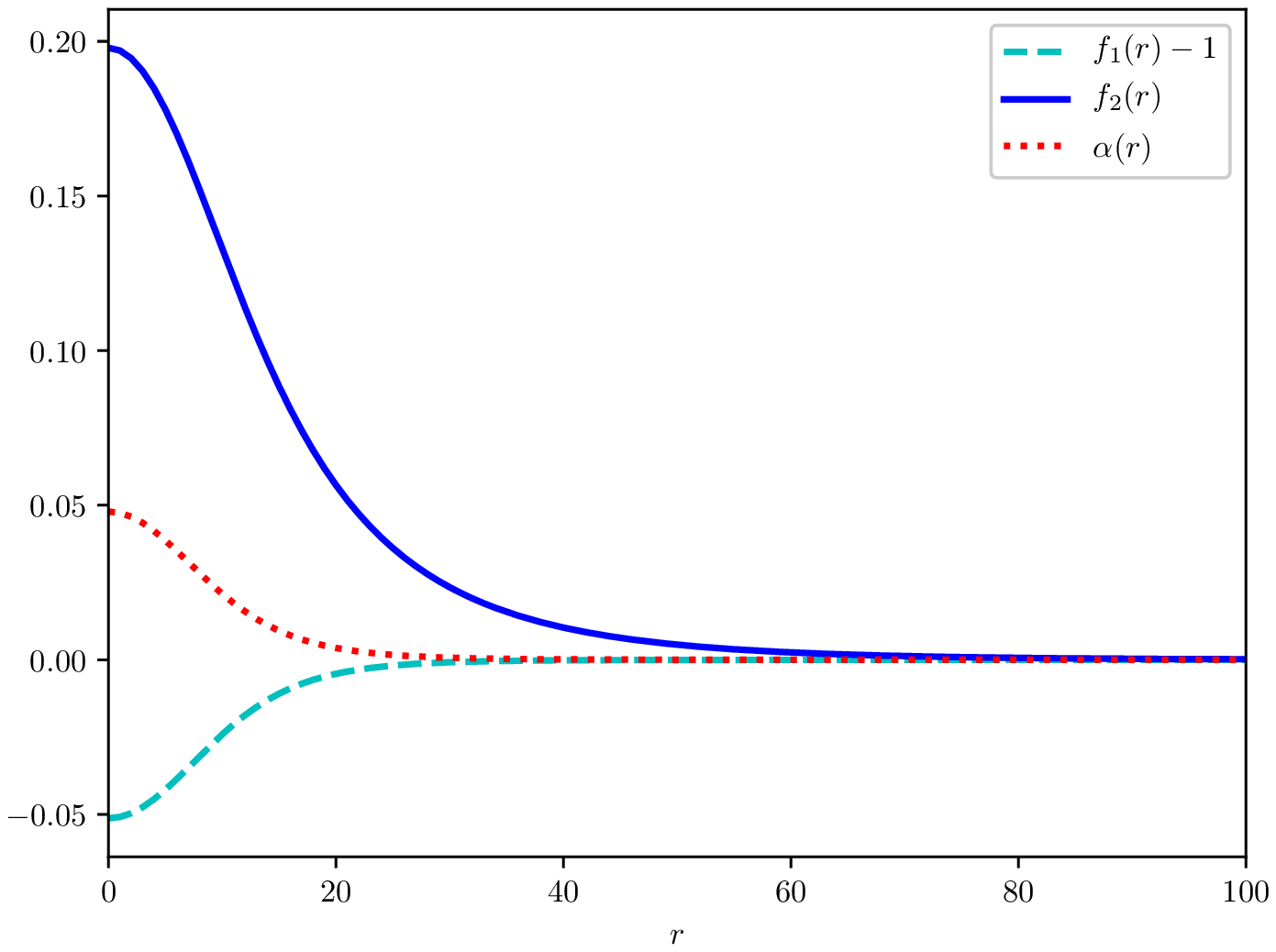}
 \caption{$\omega=1.182$}
\end{subfigure}
\begin{subfigure}[t]{0.5\textwidth}
 \noindent\hfil\includegraphics[scale=.5]{omu4.eps}
 \caption{$\omega=1.183$}
\end{subfigure}
\caption{The frequency dependence of the solutions approaching the upper frequency limit, $\beta_1=0.5$, $\beta_{12}=\mu=1.4$, $\beta_2=0$.}
\label{fig:4omu}
\end{figure}

\end{document}